\RequirePackage{fix-cm}
\documentclass[smallextended]{svjour3}       
\smartqed  

\usepackage{lscape}
\usepackage{graphicx}
\usepackage[export]{adjustbox}
\usepackage{ulem}
\usepackage{amsmath}
\usepackage{algpseudocode}
\usepackage{algorithm}
\usepackage{multirow}
\usepackage{graphicx}
\usepackage[hyphens]{url}
\usepackage[colorlinks=true, linkcolor=blue, citecolor=blue, bookmarks=true,pagebackref=false]{hyperref}
\usepackage{hypernat}
\usepackage{listings}
\usepackage{framed}
\usepackage{booktabs}
\usepackage{balance}
\usepackage{fancybox}
\usepackage{verbatim}
\usepackage[usenames,dvipsnames]{color}
\usepackage{caption}
\usepackage{rotating}
\usepackage{threeparttable}
\usepackage{amssymb}
\usepackage{pifont}
\newcommand{\cmark}{\ding{51}}%
\usepackage{longtable}
\usepackage{soul}

\usepackage{colortbl}
\usepackage{arydshln}
\usepackage{multirow}
\usepackage{multicol}

\usepackage{cite}
\usepackage{array}
\usepackage{rotating}

\usepackage{subfigure}

\newcolumntype{^}{>{\currentrowstyle}}

\newcolumntype{R}{>{\raggedright\arraybackslash}p{1cm}}
\newcolumntype{L}{>{\raggedleft\arraybackslash}p{1cm}}

\usepackage{titlesec}
\titleformat{\paragraph}
{\normalfont\normalsize}{\theparagraph}{1em}{}
\titlespacing*{\paragraph}
{0pt}{3.25ex plus 1ex minus .2ex}{1.5ex plus .2ex}

\begin{document}
\setstcolor{red}
\title{Empirical Study on the Software Engineering Practices in Open Source ML Package Repositories}



\author{Minke Xiu \and Ellis E. Eghan \and Zhen Ming (Jack) Jiang \and Bram Adams 
}


\institute{M. Xiu\at
        The Software Construction, AnaLytics and Evaluation (SCALE) Lab \\
        York University, Toronto, ON, Canada\\
        \email{xmk233@eecs.yorku.ca}
 		\and
 		E.E. Eghan\at
 		Lab on Maintenance, Construction and Intelligence of Software (MCIS) \\
 		École Polytechnique de Montréal, Montréal, QC, Canada \\
 		\email{ellis.eghan@polymtl.ca}
        \and
        Z.M. Jiang\at
        The Software Construction, AnaLytics and Evaluation (SCALE) Lab \\
        York University, Toronto, ON, Canada \\
        \email{zmjiang@cse.yorku.ca}
		\and
		B. Adams\at
		Lab on Maintenance, Construction and Intelligence of Software (MCIS) \\
		Queen’s University, Kingston, ON, Canada \\
		\email{bram.adams@queensu.ca}
}

\date{Received: date / Accepted: date}

\maketitle
	
\begin{abstract}

Recent advances in Artificial Intelligence (AI), especially in Machine Learning (ML), have introduced various practical applications (e.g., virtual personal assistants and autonomous cars) that enhance the experience of everyday users. However, modern ML technologies like Deep Learning require considerable technical expertise and resources to develop, train and deploy such models, making effective reuse of the ML models a necessity. Such discovery and reuse by practitioners and researchers is being addressed by public ML package repositories, which bundle up pre-trained models into packages for publication. Since such repositories are a recent phenomenon, there is no empirical data on their current state and challenges. Hence, this paper conducts an exploratory study that analyzes the structure and contents of two popular ML package repositories, TFHub and PyTorch Hub, comparing their information elements (features and policies), package organization, package manager functionalities and usage contexts against popular software package repositories (npm, PyPI, and CRAN). Through these studies, we have identified unique SE practices and challenges for sharing ML packages. These findings and implications would be useful for data scientists, researchers and software developers who intend to use these shared ML packages. 

\end{abstract}



\section{Introduction}
\label{sec:introduction}

The development of AI software is in great demand; AI is making revolutionary changes in healthcare, retail, energy, software development and other fields~\cite{industrieschangedbyai}. 
Deep learning, especially, shines in the fields that were originally considered to be science fiction, such as autonomous driving and voice assistants. 
Recent estimates show that Machine Learning (ML) applications have the potential to create between \$3.5 and \$5.8 trillion in value annually~\cite{MckinseyReport}. 

Unfortunately, the development of ML models, which is the core of AI software development, is not a trivial task. 
First, the implementation of ML algorithms is difficult as specific skills are required in reading and understanding professional AI literature. 
The emergence of ML frameworks (e.g., TensorFlow~\cite{TensorFlow}, PyTorch~\cite{PyTorch}, etc.), however, has greatly lowered the skill requirements for model development. 
The development of ML models with such frameworks involves calling appropriate APIs, allowing developers to devote more time and effort on other tasks such as obtaining, pre-processing, labeling and filtering data, adjusting and testing the models' structure, or proposing new deep learning algorithms. 

Secondly, training ML models requires significant resources. 
Models that perform complex tasks like image classification or text embedding need intensive calculation and require a long time to finish training on large-scale datasets~\cite{ZophCVPR2018}, which leads to substantial computational resource consumption. 
Although expensive equipment such as GPUs can be used to shorten the training duration, they may not always be available to software developers. 

As a result, there is such a high demand for shareable and reusable pre-trained models, which has resulted in the creation of so-called ML model repositories. 
Such repositories aim to bridge the gap between \textbf{AI experts} and general \textbf{users} such as (1) data scientists, (2) software engineers who wish to reuse the model and (3) researchers interested in studying the challenges and opportunities of reusing pre-trained models. 

ML model repositories can be divided into two categories based on their distribution methods.
The first category of the ML model repositories, referred to as \textbf{ML model stores}, provides cloud-based model deployment support. 
Such repositories usually come at a cost as users are required to cover the fees for hosting and using the provided cloud computing resources. 
Common examples of such model stores are AWS marketplace~\cite{awsmarketplaceformlai}, ModelDepot~\cite{modeldepot}, and the Wolfram neural net repository~\cite{wolframneuralnetrepository}. 
These are comparable to the traditional mobile app stores (e.g., Google Play~\cite{googleplay} and Apple's app store~\cite{appstore}). 
In our previous work~\cite{xiu2020exploratory}, we empirically studied model stores. 
We found that although there are notable commonalities and differences with their counterpart in traditional software development (mobile app stores), such repositories are still in their infancy. 

The second category of  ML model repositories are referred to as \textbf{ML package repositories}. 
ML package repositories contain tens to hundreds of pre-trained models specially bundled up into \textbf{ML packages} that are distributed via \textbf{ML package managers} (including ML frameworks, ML-related libraries and APIs). 
For example, the PyTorch Hub repository~\cite{PyTorchHub} contains packages that can be accessed by a user via the PyTorch framework APIs.
These ML packages are free to use, but users must manually manage and deploy these models and their dependencies. 
The most popular examples of such repositories are TFHub~\cite{tfhub} and PyTorch Hub\cite{PyTorchHub}. Their distribution practices are similar to programming language-specific software package repositories like npm~\cite{npmpage}, PyPI~\cite{pypipage} and CRAN~\cite{cranpage}.

Unfortunately, there is still no empirical data about the state-of-the-art best practices and challenges involving free ML package repositories. 
In particular, what information about packages/models is provided by ML repositories and how helpful are they to the users? 
What task types are supported by these ML packages? 
How are ML packages/models organized and distributed? 
How can users reuse such ML packages/models? 
Are there any common practices between these free ML package repositories and their counterpart software package repositories? 


Hence, in this paper we conduct an exploratory study on two ML package repositories, TFHub and PyTorch Hub. 
We not only compare the structure and information elements (features and policies) among these ML package repositories with each other, but also compare them against their counterpart software package repositories (npm, PyPI and CRAN). 
Our results show a number of significant differences between the practices of reusing ML packages and traditional software packages. 
First, we observe that although a few of the practices of software package repositories have been adopted by their ML counterparts 
(e.g., product line architecture, multiple usage contexts), most of the established SE practices are either not adopted or are in their infant stages within the ML package repositories (e.g., release management, dependency management, security, package management functionalities). 
Secondly, we observe some practices within the ML package repositories that are not yet adopted within the software package repositories (e.g., quality evaluation of packages). 

Also, the process of using ML packages differs from software packages. ML packages are individual artifacts that can be downloaded, loaded, and used in multiple usage contexts with the help of special ML framework APIs. However, traditional software packages have different downloading, loading, and usage practices from ML packages (e.g., ML packages are not installed; rather, a specific model within the package is loaded at run-time based on the arguments provided by the user). 
The findings of this research will help users (like software engineers) that are familiar with traditional software package repositories, but not yet with ML package repositories, to have a clear and easier understanding of ML package repositories. 
	
The contributions of this paper are as follows:
\begin{itemize}
 
\item We provide an overview of the current practices on sharing reusable ML packages through a study of the structure and contents of ML package repositories. To the best of our knowledge, this is the first empirical study on ML package repositories. 

\item By comparing against the sharing mechanism of software package repository, we have identified a set of unique practices and challenges on distributing, sharing, and using pre-trained ML packages. 

\item Our comparison between the practices within ML and software package repositories presents stakeholders of these repositories with opportunities of how to adopt the established practices from their counterpart repositories.


\end{itemize}

The remainder of the paper is organized as follows. 
Section~\ref{sec:background} provides background on ML package repositories and software package repositories. 
Section~\ref{casestudysetup} introduces details of our case study setups. 
We study the structure and contents of ML package repositories in Section~\ref{sec:rq1} and Section~\ref{sec:rq2}, respectively.
In Section~\ref{sec:rq3}, we investigate the functionalities and usage contexts provided by {ML package managers. 
Section~\ref{sec:relatedWorks} surveys related work. 
Section~\ref{sec:threatToValidity} explains the threats to the validity of our work. 
Section~\ref{sec:conclusions} presents the conclusions and avenues for future work. 
\section{Background} 
\label{sec:background}

\label{sec:caseStudySetup_hubs}

\begin{table}[]
\centering
\caption{Comparison of statistics of software package repositories and ML package repositories. ``+'' indicates a repository containing a small percentage of packages in other programming languages. The software package repository statistics are gathered on April 19, 2020 from \texttt{Libraries.io}. The ML model repositories' statistics are gathered on March 25, 2020.} 
\label{tab:repometa}
\begin{tabular}{|l|l|l|l|l|}
\hline
\multicolumn{2}{|c|}{\textbf{Repository}} & \multicolumn{1}{c|}{\textbf{\begin{tabular}[c]{@{}l@{}}Language(s) or\\Framework \end{tabular}}} & \multicolumn{1}{c|}{\textbf{Launch Time}} & \multicolumn{1}{c|}{\textbf{\# Packages}} \\ \hline
\multirow{15}{*}{\textbf{\begin{tabular}[c]{@{}l@{}}Software\\ Package\\ Repository\end{tabular}}} 
 &  Bower & JavaScript+ & Sep 2012 & 69,678  \\ \cline{2-5} 
 &  Cargo (Crates.io) & Rust+ & Jun 2014 & 40,142  \\ \cline{2-5} 
 &  Clojars & Clojure+ & Nov 2009 & 25,913  \\ \cline{2-5} 
 &  CRAN & R & $\ge$ Aug 1993 & 17,370  \\ \cline{2-5} 
 &  Go Package Community & Go+ & $\ge$ Mar 2008 & 1,818,628 \\ \cline{2-5} 
 &  Hackage & Haskell+ & \begin{tabular}[c]{@{}c@{}} Jun 2008 \end{tabular} & 14,758  \\ \cline{2-5} 
 &  Hex & Elixir+ & Dec 2013 & 9,911  \\ \cline{2-5} 
 &  Maven & Java+ & Sep 2003 & 185,402  \\ \cline{2-5} 
 &  MELPA (Emacs) & Emacs Lisp+ & Oct 2011 & 5,026 \\ \cline{2-5} 
 &  MetaCPAN (CPAN Search) & Perl+ & Nov 2010 & 37,790  \\ \cline{2-5} 
 &  npm & JavaScript & Sep 2009 & 1,366,638 \\ \cline{2-5} 
 &  NuGet & C\#+ & Jan 2011 & 201,192  \\ \cline{2-5} 
 &  Packagist & PHP+ & Apr 2011 & 328,953  \\ \cline{2-5} 
 &  PyPI & Python & Oct 2008 & 250,533  \\ \cline{2-5} 
 &  Rubygems & Ruby+ & Nov 2003 & 164,749  \\ \hline

\multirow{7}{*}{\textbf{\begin{tabular}[c]{@{}l@{}}ML Package/Model\\ Repository\end{tabular}}} 
 & \begin{tabular}[c]{@{}l@{}}AIHub\\TensorFlow Module\end{tabular}
 			& \begin{tabular}[c]{@{}l@{}}TensorFlow \\ (Python)\end{tabular}
 			& (March 2019) 
 			& 322  \\ \cline{2-5} 
 & DL4J Zoo Models  
  			& \begin{tabular}[c]{@{}l@{}}DL4J (Java) \end{tabular}
  			& Jun 2019 
  			& 16  \\ \cline{2-5} 

 & MXNet GluonCV Model Zoo 
 			& \begin{tabular}[c]{@{}l@{}}MXNet \\ (Python, Scala, etc.) \end{tabular}
 			& $\ge$ Apr 2014
 			& 323  \\ \cline{2-5} 
 & MXNet GluonNLP Model Zoo 
  			& \begin{tabular}[c]{@{}l@{}} MXNet (Python, Scala, etc.) \end{tabular}
  			& $\ge$ Apr 2014
  			& 42  \\ \cline{2-5} 
 & PyTorch Hub 
 			& \begin{tabular}[c]{@{}l@{}} PyTorch \\(Python, C++, Java) \end{tabular}
 			& Jun 2019 
 			& 26  \\ \cline{2-5} 
 & spaCy Models 
 			& \begin{tabular}[c]{@{}l@{}}spaCy \\ (Python) \end{tabular}
 			& $\ge$ Feb 2015 
 			& 6  \\ \cline{2-5}  
 & TFHub  
 			& \begin{tabular}[c]{@{}l@{}}TensorFlow \\ (Python, JavaScript, \\ C++, Java, etc.)\end{tabular} 
 			& March 2018 
 			& 471  \\ \cline{2-5}
 & Torch7 Model Zoo 
 			& \begin{tabular}[c]{@{}l@{}} Torch7 (LuaJIT, C) \end{tabular}
 			& Jan 2015 
 			& 20  \\ \hline
\end{tabular}
\end{table}

Package managers are a set of software tools that automate the process of package installation, upgrade, 
and removal in a consistent manner. Packages are hosted in and downloaded from package repositories$\footnote{https://en.wikipedia.org/wiki/Package\_manager}^,\footnote{https://en.wikipedia.org/wiki/List\_of\_software\_package\_management\_systems\#Application\-level\_package\_managers}$. Generally speaking, package managers can be divided into two groups.  The first group (e.g., dpkg for Debian, Homebrew for macOS, and Windows Store for Windows) provides compiled (binary) or source code package management for operating system-specific applications, while the second group (e.g., npm for JavaScript and PyPI for Python) provides package management for programming language-specific API-level packages. This research focuses only on API-level software package repositories (referred from this point simply as software package repositories), and contrasts them to ML package repositories. 

In order to gain some basic knowledge of the existing software package repositories, Table~\ref{tab:repometa} presents the basic statistics of the popular software package repositories (containing more than 4,000 packages) from \texttt{Libraries.io}~\cite{librariesio}, which is a popular index of the most common software package repositories --- it monitors the information about the packages within different software package repositories. 
As shown in the table, we measured the launch time of repositories as the time of the first commit of the GitHub repository that stores the actual source code powering the package sharing websites. There are a few exceptions: the launch times of the Go and CRAN repositories cannot be found, so we note the time as no earlier than ($\ge$) the initial date of their respective programming language (Go and R); NuGet's launch time is gathered from Microsoft's documentation, not GitHub. 

Unfortunately, since the concept of ML package repository is relatively new, their information is not tracked yet by \texttt{Libraries.io} and we have to manually gather information for ML package repositories. The list of ML package repositories is obtained from the study of Braiek et al.~\cite{BraiekMSR2018}. 
Their launch times are retrieved from multiple sources (e.g., twitter and blog posts) that announced the launch, the first release, or commit of their respective GitHub repositories. We note the launch time as no earlier than the release time of the frameworks if no information is found from the previously mentioned sources. 
The total number of ML packages are counted manually since most of the ML package repositories (except for AIHub TensorFlow module~\cite{aihubtfmodule} and TFHub~\cite{tfhub}) do not provide these statistics. The results are shown in Table~\ref{tab:repometa}. 

Due to their recency and relatively high learning curve (requiring deep ML expertise) and computing resource requirements, there are fewer reusable packages in ML package repositories, compared to software package repositories. We observe an average growth of approximately eight new packages per week in TFHub and two new packages every week in the AIHub repository, indicating that ML package repositories are still an upcoming phenomenon. 
Despite the relatively slow growth of the number of packages within these ML repositories, we observe a high usage of TFHub and PyTorch Hub ML packages in open source projects. 
For example, a preliminary search of PyTorch Hub package loading API keyword$\footnote{``\texttt{https://github.com/search?q=torch.hub.load\%28\&type=Code}''}$ shows that PyTorch Hub packages are loaded in over 143K source code files on GitHub. Another search$\footnote{``\texttt{https://github.com/search?q=torch.hub.load(`huggingface/transformers'\%28\&type=Code}''}$  showed that Huggingface Transformers, a package containing most of the complex NLP models, is used within 3.2K source code files on GitHub.
These results show that users prefer to reuse such existing models given the difficulty of training custom ML packages. Thus, these limited number of ML packages can power unlimited possibilities in ML software development and ML research.

\section{Case Study Setup}
\label{casestudysetup}

This paper is an exploratory study of the ML package repositories addressing the following research questions (RQs).

\textbf{RQ1 - What types of information are presented for software package repositories and ML package repositories?} 
Given the relatively short existence of ML package repositories, this RQ (Section~\ref{sec:rq1}) aims to provide us with insights on the structure (e.g., the organization of packages by task types) and practices (e.g., release management) of the ML repositories, as well as to discover any missing or non-formalized information elements (based on the comparison with their counterparts in the software engineering domain). 
Such discoveries will be helpful in building a better ML package repository in the future in the sense of providing more information transparency, benefiting more users, especially software engineers without solid background in ML. 

\textbf{RQ2: How are packages organized in ML package repositories?}  Next, we investigate  the organization practices of ML package repositories, in Section~\ref{sec:rq2}. More specifically, we study the family phenomenon within these repositories, and its implications on package task type distribution, package similarity, and release management. 


\textbf{RQ3 - What is the process needed in order to use the functionalities from software/ML package repositories?}
Finally, we study the functionalities of the package managers (tools and libraries) provided by the ML package repositories and how they are used in Section~\ref{sec:rq3}. These ML package managers, as mentioned in the introduction of this paper, provide functionalities that allow users to explore, manage and uses the ML packages. In this RQ, we aim to discover the unique practices of the ML package managers, and compare them with the practices of traditional software package managers. The findings and implications of this RQ may point out if there are any practices of traditional software package managers that can be adopted to improve the functionality of ML package managers. 

For each RQ, the first section describes our approach, the second section presents our findings, followed by the third section discussing the results and their implications. However, before addressing the RQs, we first discuss the five repositories selected from Table~\ref{tab:repometa} as the subjects of our exploratory study. 

Among the studied ML package and software package repositories mentioned in Section~\ref{sec:background}, our case study will focus on the npm, PyPI and CRAN software package repositories, and the TFHub and PyTorch Hub ML package repositories. The rationale of selecting these repositories is as follows:

\begin{itemize}
	\item npm, PyPI, and CRAN respectively are the top three ``mono-language" software package repositories in terms of the number of packages they host, based on the statistics presented in Table~\ref{tab:repometa}. They cover JavaScript, Python and R, respectively
	
	\item TFHub and PyTorch Hub are the official repositories of TensorFlow and PyTorch, the most popular ML frameworks in academia and industry~\cite{mostwidelyusedframework,pytorchgettingattention}. 	
\end{itemize}

\section{RQ1: What types of information are presented for software package repositories and ML package repositories?} 
\label{sec:rq1}

In this RQ, we aim to understand the types of information presented in the software package and ML package repositories. We focus on the information elements (IE), each of which describes one aspect of the packages or the repository, e.g., the basic description of the package, the dependencies of the package, parameter value settings, etc. 
By comparing the IEs in ML and software package repositories, we learn of the structure of the repositories, as well as the missing and new/additional IEs needed by ML packages and ML package repositories. In what follows, we present the methodology used to achieve this (based on our earlier work~\cite{xiu2020exploratory}), and a discussion of our findings.


\subsection{Approach}

Here we explain our process of extracting IEs from different repositories. 
This process is based on the process that has been used in our earlier work on ML model stores~\cite{xiu2020exploratory}. 

\begin{figure} 
	\centering 
	\includegraphics[scale=0.3]{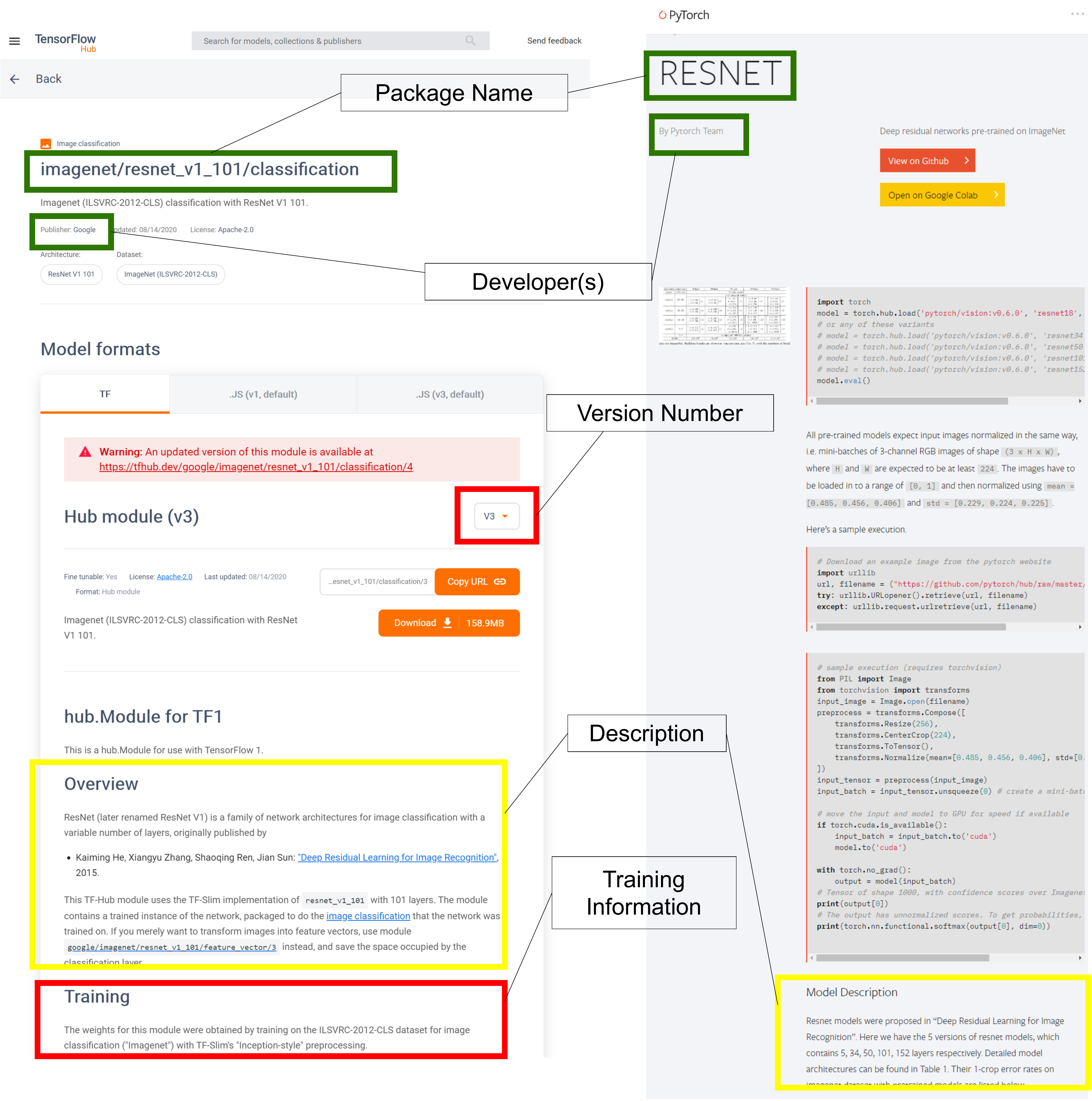} 
	\caption{An example comparing the IEs in TFHub (left) and PyTorch Hub (right) packages. The example highlights the same IE in both repositories (green), similar IEs with different representations (yellow), and IEs unique to only one repository (red)} 
	\label{fig:IE_Example} 
\end{figure}

We refer to several sources to determine the full list of IEs per repository. The fundamental source is the website of the packages. Secondary sources such as the software package description/documentation files$\footnote{https://docs.npmjs.com/files/package.json}^,\footnote{https://github.com/pypa/sampleproject/blob/master/setup.py}^,\footnote{https://cran.r-project.org/web/packages/policies.html\#Source-packages}$, ML package contribution instructions$\footnote{https://github.com/tensorflow/hub/blob/master/tfhub\_dev/README.md}^,\footnote{https://github.com/pytorch/hub/blob/master/docs/template.md}$ and the JSON data structures generated upon loading a package website were also analyzed. 

The first author independently analyzed the information sources in a first iteration to identify 39 IEs.  Following several discussions and 6 more iterations of analysis involving three of the authors, a final set of 33 IEs  was agreed upon. The first iteration of analysis by the authors achieved an IRR score (based on the Cohen Kappa coefficient~\cite{landis1977measurement}) of 94.87\%, and increased to 100\% at the last iteration. 

It is normal for repositories to present the same IE differently, especially by referring to the same IE by different terms. For example, Figure~\ref{fig:IE_Example} shows how IEs are presented on sample packages on the TFHub and PyTorch Hub repositories.
In our approach, we manually unify the IEs that are essentially the same. Furthermore, we group related IEs into \textbf{dimensions}. 
We refer to the previous work of Bommarito et al~\cite{bommarito2019empirical} to verify the sanity of IEs. In their work, the authors used several IEs extracted from PyPI to study the repository's evolution and the distributions of important statistics (package release numbers, authors' package numbers, package import numbers, sizes, etc.). 
The IEs (packages, releases, dependencies, category classifications, licenses, package imports, authors, maintainers, and organizations) investigated by Bommarito et al. can all be extracted from our analyzed repositories, except for package imports, which is an internal property of packages and not related to this RQ. 

The final list of extracted IEs are presented in Table~\ref{tab:mergedTable1}. Each row is an IE, and the \cmark~represents its presence in the associated repository$\footnote{The IEs are extracted from software package and ML package repository pages at the end of March, 2020.}$. 
For example, in the \textbf{Package Information} dimension, the \textbf{Demo} row shows that npm, TFHub and PyTorch Hub have this IE while PyPI and CRAN do not.

IE dimensions are ordered alphabetically, and the IEs in every dimension are implicitly grouped as follows. 
The first group contains IEs that belong to all five repositories. 
The second group contains IEs that exist only in software package repositories (at least one). 
The third group contains IEs that exist only in ML package repositories (at least one). 
The fourth group contains other IEs in the dimension that cannot be classified into the aforementioned groups, i.e., IEs that belong to some of the ML repositories and some of the software repositories}. Within each implicit group, the IEs are ordered alphabetically. 

\begin{landscape}
\begin{table*}[]
    \centering
    \caption{Information elements among the three software package repositories and two ML package repositories. Information elements in bold are unique to either software package repositories (at least one) only or ML package repositories (at least one) only.} 
    \label{tab:mergedTable1}
    \begin{threeparttable}
    \scalebox{0.7}{
	\begin{tabular}{p{3cm}p{3cm}p{0.7cm}p{0.7cm}p{1cm}p{1cm}p{1.3cm}p{12cm}}
    \toprule
    \multirow{2}{*}{\textbf{Dimension}} & \multirow{2}{*}{\textbf{Element}} & \multicolumn{3}{c}{\textbf{SW Pkg Repo}} & \multicolumn{2}{c}{\textbf{ML Pkg Repo}} & \multicolumn{1}{c}{\textbf{Description}} \\ 

    & & (\textbf{npm} & \textbf{PyPI} & \textbf{CRAN}) & (\textbf{TFHub} &\textbf{PyTorch Hub}) & \\ 
    
       	\midrule
       	
       	\multirow{4}{2cm}{\centering Delivery} 
       	
       	& Dependencies & \cmark & \cmark & \cmark & \cmark & \cmark & The packages this package depends on. Without them, this package cannot function normally. \\ 
       		
       	& Running Environment & \cmark & \cmark & \cmark & \cmark & \cmark  & Information related to usage context. E.g., multiple formats, hardware and software environment requirements. \\ 
       	\cdashline{2-8}[0.8pt/2pt]
       	
       	& \textbf{Dependents} & \cmark &  & \cmark &  &  & The packages that depend on this package. Without this package, those packages cannot function normally. \\ 
       	
       	\cdashline{2-8}[0.8pt/2pt]
       	
       	\cdashline{2-8}[0.8pt/2pt]
    	& Downloadable Provided &  & \cmark & \cmark & \cmark &  & The package can be downloaded as files or a zipped file other than using deployment, installation or initialization command \\ 
    
       	\midrule
       	\multirow{2}{3cm}{\centering Legal Information} 
       	
       	& License & \cmark & \cmark & \cmark & \cmark &  & Regulations on how users can use this package.  \\ 
    	\cdashline{2-8}[0.8pt/2pt]
       	& \textbf{Copyright} &  &  & \cmark &  &  & Information claiming the owner of the copyright (not necessarily the author). \\ 
    
    \midrule

    \multirow{12}{2cm}{\centering Package Information} 
	
	& Description & \cmark & \cmark & \cmark & \cmark & \cmark & The objectives, functionalities and other basic information of this package. \\ 
		 
		 & Developer(s) & \cmark & \cmark & \cmark & \cmark & \cmark & Owner, publisher or collaborators of this package.  \\ 
		 
		 & Extra Information & \cmark & \cmark & \cmark & \cmark & \cmark & Other information related to this package, e.g., homepage links, more resources, academic references. \\ 
		 
		 & Indexing Keywords & \cmark & \cmark & \cmark & \cmark & \cmark & Keywords of this package that can be used to search similar packages. Classify and select packages according to keywords. \\ 
		 
		 & Package Name & \cmark & \cmark & \cmark & \cmark & \cmark & Name of the package. \\ 
	\cdashline{2-8}[0.8pt/2pt]
	& \textbf{\# Downloads} & \cmark &  &  &  &  & Number of downloads of this package. \\ 
		 
		 & \textbf{GitHub Statistics} & \cmark & \cmark &  &  &  & E.g., \# PRs, \# Issues, \# Stars, \# Forks. \\ 
	\cdashline{2-8}[0.8pt/2pt]
	& Demo & \cmark &  &  & \cmark & \cmark & A functionality through which users can try before downloading/deploying the package. \\ 
	 
		 & Published Time & \cmark & \cmark & \cmark & \cmark &  & The publish date of this version.  \\ 
		 
		 & Size & \cmark &  &  & \cmark &  & Package's disk size.  \\ 
		 
		 & Version Alert & \cmark & \cmark &  & \cmark &  & Information indicating that this package is either not the latest version or not suitable for use. \\ 
		
		 & Version Number & \cmark & \cmark & \cmark & \cmark &  & The version number of the current release.  \\ 
	 
	 	   	\midrule
	 	   	
	 	   	\multirow{2}{3cm}{\centering  Package Submission \& Review 
	 	   	} 
	 	   	
	 	   	& Developer Contribution & \cmark &\cmark & \cmark & \cmark & \cmark & General developers have access to submit and publish packages.\\ 
	 	   	
	 	   	& Review Mechanism &  &  & \cmark & \cmark & \cmark & Review process for submitted package. E.g., policies, pull request review. 
	 	  	 \\ 
	 
	   	\midrule
	   	
	   	\multirow{1}{2cm}{\centering Security} 	
	   	
	   	& \textbf{Vulnerability Report} & \cmark &  &  &  &  & Report a security vulnerability for this package. \\ 
	 
	 \midrule
	 
	 \multirow{2}{3cm}{\centering Software Development} 
	 
	 & Issue Tracking Information & \cmark & \cmark & \cmark & \cmark & \cmark & Information for reporting issues. E.g., GitHub repository issue link. \\ 
	 
	 & Source Code Repo & \cmark & \cmark & \cmark & \cmark & \cmark & The GitHub or other repository for this package.  \\ 

    \midrule 

    \multirow{11}{2cm}{\centering Technical Documentation}  
    
    & User Instruction & \cmark & \cmark & \cmark & \cmark & \cmark & Instructions on how to use this package. E.g., example code snippets. \\ 
    \cdashline{2-8}[0.8pt/2pt]
    & \textbf{Package Component Information} & \cmark & \cmark & \cmark &  &  & Information about the package's components. E.g., source code file location, data file location.    \\ 
    \cdashline{2-8}[0.8pt/2pt]
    & \textbf{Algorithm} &  &  &  & \cmark & \cmark & Information describing the algorithm of this package. E.g., neural network architecture. \\ 
        
	    & \textbf{Data Description} &  &  &  & \cmark & \cmark & Information about the dataset. E.g., dataset name, IO data shape, data pre-procession. \\ 
	    
	    & \textbf{Package Quality Evaluation} &  &  &  & \cmark & \cmark & Information about how good the package is. \\ 
	    
	    & \textbf{Training Information} &  &  &  & \cmark &  & Details about training this package. E.g., the training checkpoint file used, hyper-parameter settings. \\ 
    \cdashline{2-8}[0.8pt/2pt]
    & Pre-defined Interfaces & \cmark & \cmark &  & \cmark & \cmark & 
      E.g., the entrypoint(s) of the program, the main file/ function of a package, a special function related to a command. \\ 
    
	    & Package Domain & \cmark & \cmark &  & \cmark & \cmark & The application domain or task type of the package, e.g., \texttt{image classification} for ML packages, \texttt{front-end} for software packages.  \\ 
	    
	    & Release History & \cmark & \cmark & \cmark & \cmark &  & Accessible old versions of this package.  \\ 
	        
	    & Release Notes &  &  & \cmark & \cmark &  & Information explaining the changes that have been made for each release.  \\ 

    
   \bottomrule

    \end{tabular}
	}
\end{threeparttable}
\end{table*}
\end{landscape}

\subsection{Findings} 

\subsubsection{Comparison between Software Package Repositories and ML Package Repositories}
\label{sec:rq1-comparebetween2typesofrepo}

We analyze the common and unique IEs between the two types of repositories: software package and ML package repositories. It should be noted that if we say a type of repository has an IE, it means that this IE appears in at least one of the repositories of this type. For example, since npm has \texttt{Demo}, we say software package repositories have this IE. By  comparing between the two types of repositories, we can identify what meaningful IEs are missing in ML package repositories. \\



In the \textbf{Delivery} dimension, both types of repositories have four IEs each, with three IEs (\texttt{Dependencies, Running Environment, Downloadable Provided}) in common. 

While software package repositories usually provide a special area for dependencies on the package's webpage, dependencies in ML package repositories are often in free-text within the package description. Furthermore, ML packages' dependencies are usually Python packages that can be installed from PyPI, e.g., dependency software packages of ML packages typically are related to data processing like \texttt{opencv-python} $\footnote{https://pypi.org/project/opencv-python/}$, \texttt{tensorflow-text} $\footnote{https://pypi.org/project/tensorflow-text/}$, rather than other ML packages. 
 
Although the \texttt{Dependents} IE is unique to software package repositories, we found evidence of its possible inclusion in ML packagage repositories given the observed dependencies between ML packages such as \texttt{llr-pretrain-adv-latents} and \texttt{llr-pretrain-adv-linear}. The output of the former is the input of the latter, and the output of the latter ML package can be used as the basis for classification. In other words, to complete the classification task, two ML packages must be used in combination. PyTorch Hub also has similar examples. The output of ML package \texttt{Tacotron2} is used as the input of ML package \texttt{WaveGlow}. The two ML packages can be used together to complete the text to speech task. Because the existence of such examples, it is worthy to consider adding a special dependents/dependencies IE to the ML package page to illustrate the interrelationship between ML packages. 



In the \textbf{Legal Information} dimension, \texttt{License} is the only common IE between both types of repositories. Packages on both type of the repositories are licensed. Software package developers need to explicitly specify a license in the description file of the software package. In ML package repositories, however, packages bear the default license unless specially declared. 

If the \texttt{Copyright} belongs to people other than the author, the copyright holder is also needed to be specified. This IE is only found in the CRAN software package repository.\\


In \textbf{Package Information}, Software package repositories have 12 IEs, including all ten IEs that ML package repositories have. 

The \texttt{Developer}s of software packages tend to be individuals while the ML package owners are usually organizations. Generally, software packages are less likely to be connected with an organization. Previous studies show that 75\% of npm packages are published by individual developers~\cite{lertwittayatrai2017extracting}, and only about 5\% of PyPI package authors are organizations~\cite{bommarito2019empirical}. Conversely, we observe that both the TFHub and PyTorch Hub repositories each contain only three ML packages published by individual developers.



There is no formal versioning mechanism in ML package repositories. Software package repositories mostly adopt the semantic versioning format, e.g., \texttt{x.y.z} where \texttt{x} means a major change, \texttt{y} means a minor change and \texttt{z} means a patch~\cite{cogo2019empirical}. In TFHub, the version number is simply represented by integers like version \texttt{1}, \texttt{2} and \texttt{3}. However, there is no versioning mechanism in PyTorch Hub. 

\texttt{\# Downloads} and \texttt{GitHub Statistics} are indicators of the popularity of packages, which are completely missing in ML package repositories. 
Such an IE can reflect the wide usage of package and its huge possibility of satisfying the need of most developers. This will be helpful for ML package users, especially for people without solid ML expertise (like general software engineers), but might not be of much use to experienced data scientists and ML researchers. \\


All the IEs in the \textbf{Package Submission \& Review} dimension are found in both types of repositories. However, different mechanisms are used in both types of repositories --- software package repositories have special command line tools and review processes while ML package repositories' contribution is based on GitHub pull request. For example, npm and PyPI users use command line tools to submit their locally developed projects to the package repositories. Successfully submitted packages do not undergo any further review and are made immediately available for other users (except for CRAN, which has specific contribution policies$\footnote{https://cran.r-project.org/web/packages/policies.html}$). 
In contrast, the contribution of ML package repositories is based on GitHub pull requests. 
Everyone can contribute their ML packages to TFHub and PyTorch Hub by creating a pull-request. 
Pull request-based submission mechanism can be a good point for ML packages, since pull requests usually go through a specific review process before they are merged into code base and afterwards being available to users.\\


In the \textbf{Security} dimension, only the npm package repository provides the \texttt{Vulnerability Report}. Though ML packages may also suffer from vulnerability issues, no such IE is currently provided in ML package repositories. 
According to Wang et al.~\cite{wang2018great}, attackers can use the publicly available knowledge (algorithms, dataset, architecture, etc.) of pre-trained models to create vulnerabilities to undermine the performance of dependent models (models built based on existing pre-trained models). In their work, Wang et al. demonstrated how vulnerabilities can be created for models pre-trained on the ImageNet~\cite{deng2009imagenet} dataset using the VGG~\cite{simonyan2014very} and ResNet~\cite{he2016deep} algorithms. 
Due to the existence of similar packages/models on TFHub and PyTorch Hub, it is extremely important for ML repositories to provide users with vulnerability detection and reporting functionalities.
\\


In the \textbf{Software Development} dimension. ML package repositories and software package repositories share all of the IEs. 
Software package repositories provide the GitHub Issue link of a package for users to report issues. This 
practice is different from the \texttt{Issue Tracking Information} of ML package repositories (described further in Subsection~\ref{sec:rq1sub1}). 
All of the software package repositories, just like PyTorch Hub, provide the \texttt{Source Code Repository} GitHub links of software packages/ ML packages in a fixed area on the package's page. \\


In \textbf{Technical Documentation}, software package repositories have six IEs while ML package repositories have nine. 
Among those IEs, \texttt{User Instruction, Pre-defined Interfaces, Package Domain, Release History, Release Notes} are common in both types of repositories. 	

The \texttt{Release Notes} are organized differently in the different types of repositories. For example, CRAN's release note can either be a GitHub commit message or an HTML page containing all the correlated information. In TFHub, the release notes will be directly presented at the end of the description and they generally follow a certain format. 


ML package repositories have four unique IEs.  
The IEs \texttt{Algorithm, Data Description, Package Quality Evaluation, Training Information} are all ML related and need a certain amount of ML expertise. Such IEs provide transparency to ML package users. Unlike ML packages, most software packages are not developed based on a single particular algorithm or dataset. Thus, their implementation details are rarely a concern for users; users just access the functionality of the package through its provided APIs.

While software packages are considered of high technical quality based on the proportion of passed test cases, ML packages' technical quality is determined by statistical evaluation criteria. This indicator is not in a formalized structure across ML package repositories (e.g., there is no fixed area on an ML package's site to show its performance) and it requires the users to have a relatively good understanding of the related ML task types (e.g., ML packages in image classification task type use top-1 or top-5 accuracy to evaluate their performance, ML packages in object detection task type use intersection over union). 
In our previous study on model stores~\cite{xiu2020exploratory}, we found that limited information in terms of IEs like source code and training dataset (like how dataset are pre-processed) may introduce the hidden bias to the model re-usage.
Thus, providing more ML implementation-related information and making such information more detailed and clearly organized will help users better understand use and modification of the ML packages easier.

Software package repositories do not directly provide package quality evaluation information. The quality of a software package is reflected through its usage statistics. So software package repositories may need to add a quality evaluation IE to show the test coverage results or CI results (usually captured in the external websites of software packages) within the repository for users.\\  
\subsubsection{Comparison among ML Package Repositories}
\label{sec:rq1sub1} 

There are 28 IEs (belonging to six dimensions) in at least one of the ML package repositories. Among them, 16 are common in both of the ML package repositories. We present a detailed analysis of the results of each dimension below. By doing these comparisons between ML package repositories, we will identify the common IEs of both ML package repositories and what unique IEs of one repository should also be possessed by another one.\\


In the \textbf{Delivery} dimension, ML package repositories have two IEs in common (\texttt{Dependencies} and \texttt{Running Environment}) while the \texttt{Downloadable Provided} IE is unique to TFHub.

In TFHub, the running environment IE describes the format(s) of the package and version of TensorFlow required to load the package. TFHub packages can contain models of different formats e.g., general package formats (hub.Module, TF2 SavedModel), format for TensorFlow on JavaScript, and format for deployment on edge device (computational equipment that have limited computational resource, like mobile phone). The format of a packages/model determines its usage context (more details are in Section~\ref{rq3-deployment}). 
PyTorch Hub packages do not provide the packages that are suitable for multiple usage context. However, they use the running environment IE to indicate if an ML package needs accelerator support like GPU and CUDA\cite{whatiscuda}(NVIDIA parallel computing architecture for GPU); having an accelerator can make a big difference in ML efficiency. 

Only TFHub provides links for directly downloading the package files in multiple formats.\\


\textbf{Legal Information} contains a single IE related to licensing.  
According to the TFHub contribution tutorial, if no license is specified, the default license for an ML package will be Apache 2.0\cite{apache2license}. No information was found about the licenses used by PyTorch Hub packages. \\


\textbf{Package Information}. ML package repositories have ten IEs in this dimension, with \texttt{Description, Developer(s), Extra Information, Indexing Keywords, Package Name, Demo} are common between the two repositories. The \texttt{Description} are actually in freestyle and may contain other IEs. 
ML packages generally will provide academic papers and GitHub links as \texttt{Extra Information}. These academic papers are usually the original sources of the algorithms used by the ML packages. So this IE can be helpful for users like data scientists and researchers who may benefit from dig into the basic principle of the algorithms. 
As for the \texttt{Indexing Keywords}, they can be task types (text embedding, image classification, etc.), datasets (ImageNet, etc.), algorithms (CNN, Transformer, etc.) and some other ML attributes that help users narrow down the search scope. 
The \texttt{Name}s of TFHub ML packages are more complicated than PyTorch Hub ML packages, because the former contain not only the key algorithm names but also the names of dataset or configuration values used during training. 
Their \texttt{Demo} is supported by Google Colab, an online Python notebook environment. These example notebooks usually contain complete use cases of this ML package. 

There are four other IEs unique to TFHub. 
Due to Pytorch Hub's lack of a versioning mechanism, the \texttt{Version Alert, Version Number}, and \texttt{Publication Time} are missing. Also, there is no \texttt{Size} information in PyTorch Hub. We discuss the implication of this lack of a formalized versioning mechanism in Section~\ref{sec:rq1-findings}. \\ 




Both IEs in the \textbf{Software Development} dimension are found in the two ML package repositories. 

Regarding the \texttt{Issue Tracking Information} IE, PyTorch Hub provides a link to its GitHub issue page while TFHub provides a form for users to submit any kind of feedback directly from the repository. In both ML package repositories, users can also report bugs identified in either the frameworks or ML packages on the \texttt{tensorflow\_hub} library's GitHub repository$\footnote{https://github.com/tensorflow/hub}$ and PyTorch Hub's GitHub repository$\footnote{https://github.com/pytorch/hub}$. 

Though both repositories also provide links to the \texttt{Source Code Repository}, most packages in TFHub include these links as freestyle text in the package description (not in a dedicated area like in PyTorch Hub), making this information difficult to identify. \\ 


The \textbf{Technical Documentation} dimension has six common IEs (\texttt{User Instruction, Algorithm, Data Description, Package Quality Evaluation, Training Information, Package Domain}) in both ML package repositories. However, these common IEs are usually not organized as independent IEs or formally presented; they are a part of the ML package description.

The \texttt{Package Domain} can usually be the indexing keywords of the ML packages. It contains ML application domains (computer vision, natural language processing, etc.) and ML task type (image classification, text embedding etc.).
Note that ML package repositories do not have \texttt{Package Quality Evaluation} in all of their ML packages. Only 17 (out of 26) and 35 (out of 383) packages in the PyTorch Hub and TFHub repositories, respectively, provide their quality evaluation result. In TFHub this IE is also a part of the general package description rather than independent IE. 

In addition to the shared IEs, both ML package repositories have unique IE(s). 
Because of the lack of versioning mechanism, PyTorch Hub does not have \texttt{Release History} and \texttt{Release Note}. 
As for \texttt{Training Information}, TFHub ML packages may provide more detailed algorithm information like the model optimizer arguments$\footnote{https://tfhub.dev/deepmind/spiral/default-fluid-gansn-celebahq64-gen-19steps/1}$. PyTorch Hub usually does not provide such details, but it has a unique area in a ML package page for listing the \texttt{Pre-defined Interfaces} which are the entrypoints. Entrypoint is a mechanism in PyTorch Hub to manage variant models within a single ML package. Users specify an entrypoint when loading a PyTorch package to get the needed variant of a model. On the other hand, TFHub ML packages do not have entrypoints but rather utilize a signature mechanism. The signature mechanism is used by TFHub packages to organize combinations of input and output tensors (basic data structure in ML). These two mechanisms will be explained in detail in Section~\ref{sec:rq3}. 

We find that only 173 TFHub packages (out of 383) introduced signatures (e.g., telling users about what this signature can do, what hyper-parameter it needs) in their description section. 
Though the signatures are not listed formally, the TFHub users can get a full list of the signature they supports by calling an API of the packages. In PyTorch Hub, only 10 (out of 26) packages provide a list of their entrypoints within their package pages. \\
\subsection{Implications}
\label{sec:rq1-findings} 

\noindent
\textbf{Dependency Management}. 
ML package repositories currently assume that, unlike software packages with several dependencies and dependents, ML packages rarely depend on each other but rather depend on existing Python packages and the core ML frameworks. However, our analysis was able to identify dependencies between ML packages (as discussed in Section~\ref{sec:rq1-comparebetween2typesofrepo}).
 

Hidden dependencies are a major risk for developers. Dependency-related information helps developers to make a better estimation of the effort needed to upgrade to a given software/ML package. For example, users may be concerned of the risk of introducing bugs or breaking changes during an upgrade, as well as the extra work needed to make their current dependencies compatible with the new dependencies introduced by the included software/ML package. Thus, dependency-related IEs may have an impact on when and whether developers decide to upgrade the ML package. \\

\noindent
\textbf{Release Information}. 
TFHub implements a basic incremental versioning mechanism (version numbers and release notes) while PyTorch Hub has no mechanism in place. 
Versioning helps users learn whether the ML packages have been updated and whether it is worthy to upgrade to a new release.
ML package repositories can adopt a practice similar to the semantic versioning of software package repositories to indicate the severity of changes and backward-compatibility of APIs (more details on this discussion in versioning is provided in Section~\ref{sec:rq2}). Given the numerous points of change in ML packages (e.g., algorithm, training dataset, configurations, etc.), a consensus among ML package stakeholders would have to be reached on the definition of a major and minor changes, as well as patches.\\

\noindent
\textbf{Popularity Indicator}. Experienced users of ML packages can tell which ML package is better by looking at the performance evaluation result, while users without solid ML expertise (like general software engineers or researchers not in ML area), however, may refer to information such as the popularity, reviews and the quality of technical functionalities when deciding on a ML package or software package to use. Intuitively, such users may choose the packages with the most downloads (popularity)~\cite{evaluatepackagewithpopularity} or the most positive reviews from other users (common in some traditional software engineering repositories like mobile app stores)~\cite{vasa2012preliminary}. 

Such indicators make it easier for users who do not know how to differentiate between algorithm performance and datasets to know which ML package is the most popular or has a good reputation. So the quality indicator information elements are highly recommended to be provided by ML package repositories.\\ 

\noindent
\textbf{Security}. 
TensorFlow, PyTorch and most of the other popular ML related libraries are mainly Python-based and can be installed through PyPI. However, since PyPI does not provide any submission review and security vulnerability report mechanism, this increases the quality and security risks of ML packages. These security risks are hard to discover. For example, an ML engineer may expend much effort  to locate a bug in the model's source code, but the bug may actually originate in an imported Python library installed from PyPI. 

Although the bugs may originate from the external Python dependencies, there are several vulnerability analysis tools (e.g., WhiteSource~\cite{whitesource}, snyk~\cite{snyk}) that ML developers need to include in their workflow to identify the propagation of vulnerabilities from dependent packages into the ML application.\\
\noindent
\textbf{Technical Documentation}. 
There is no formalized (or unified) structure or organization of the technical information within ML package repositories. 
ML packages have some unique technical documentation like algorithms, dataset description, training details, ML package tune-ability and ML package quality evaluation. 
Given this lack of formalism, users of ML packages need to read the documentation or description of packages in order to extract such information. These technical documentation vary in style and form, making it difficult for users to understand and compare ML packages. 
It would be beneficial if packages within a ML package repository are required to follow a documentation standard that ensures the same structure of information elements across different packages.\\
\section{RQ2: How are packages organized in ML package repositories?}
\label{sec:rq2}

In Section~\ref{sec:rq2}, we identify and compare the IEs presented in software and ML package repositories, however this RQ is specific to the domain of machine learning (packages). In this RQ, we go one step further by investigating some of the major IEs (e.g., task types, algorithms, datasets) and the inter-relationship (e.g., distributions) of ML packages within the ML package repositories. In particular, the analysis performed in this RQ is centered around a unique phenomenon in ML package repositories: ML package/model families. These are groups of packages/models that are similar with each other in terms of task types, algorithms and datasets. Thus, through the study of this RQ, we provide details on the organization practices within such ML package repositories and provide the users of the ML package repositories information about the kind of packages to expect in each ML repository. 

\subsection{Approach} 
\label{sec:rq2-datapreparation}
Using web crawlers and custom scripts (see our replication package~\cite{replicationpackage}), we extract the IEs from the JSON data structures (generated by each ML package's individual page) and webpages of each package in the TFHub and PyTorch Hub repository.  

It should be noted that PyTorch Hub provides models in PyTorch's general formats (.pt or .pth files) only. So in order to perform a fair comparison, we only consider the two general formats of TFHub, i.e., hub.Module and TF2 SavedModel. 
For example, we do not consider TFHub packages that are not provided in either hub.Module or TF2 SavedModel formats (e.g., \texttt{mobilenet\_v2\_1.0\_224\_quantized} $\footnote{https://tfhub.dev/tensorflow/coral-model/mobilenet\_v2\_1.0\_224\_quantized/1/default/1}$). 
However, if a package provides extra formats in addition to the two general ones (e.g. \texttt{imagenet/mobilenet\_v2\_075\_96/feature\_vector}  $\footnote{https://tfhub.dev/google/imagenet/mobilenet\_v2\_075\_96/feature\_vector/4}$ ), we only take those two general formats into account. Thus, the analysis in this RQ is performed on 383 TFHub packages (741 versions) and 26 PyTorch Hub packages (including 132 models). The snapshots for both repositories were taken in the middle of March 2020.  

For each ML package, we extract information about the task type, algorithm, and training dataset. 
For TFHub, the values and contents of IEs in our research scope are extracted automatically from the JSON data structure. 
Although PyTorch Hub did not provide such a data structure, we manually extract the needed information from the PyTorch Hub packages due to their limited number. 

\subsection{Findings}
\label{sec:rq2_findings}

\subsubsection{Task Type of ML Packages/Models}
\label{sec:rq2-tasktype}

\begin{table}[]
\centering
\caption{Task types and ML package/model distribution on TFHub and PyTorch Hub}
\label{tab:tasktypeDistribution}
\begin{tabular}{|c|c|c|c|c|l|}
\hline
\multirow{2}{*}{\textbf{\begin{tabular}[c]{@{}c@{}}Application\\ Domain\end{tabular}}} & \multirow{2}{*}{\textbf{Input}} & \multirow{2}{*}{\textbf{\begin{tabular}[c]{@{}c@{}}ML \\ Task Type\end{tabular}}} & \multicolumn{2}{c|}{\textbf{\# ML Packages}} & \multicolumn{1}{c|}{\multirow{2}{*}{\textbf{Task Type Description}}} \\ \cline{4-5}
 &  &  & \textbf{\begin{tabular}[c]{@{}c@{}}TFHub \\(\%)\end{tabular}} & \textbf{\begin{tabular}[c]{@{}c@{}}PyTorch Hub \\(\%)\end{tabular}} & \multicolumn{1}{c|}{} \\ \hline

\multirow{2}{*}{\begin{tabular}[c]{@{}c@{}}Audio\\ Processing\end{tabular}} 
 & \multirow{2}{*}{Audio} & Embedding & 3 (0.8\%) & 0 (0.0\%) & Changing the audio into a mathematical vector.\\ \cline{3-6} 
 &  & \begin{tabular}[c]{@{}c@{}}Pitch \\ Extraction\end{tabular} & 1 (0.3\%) & 0 (0.0\%) & Recognize the dominant pitch in sung audio.\\ \cline{2-6} 
 
 & \begin{tabular}[c]{@{}c@{}}Mel\\ Spectrogram\end{tabular} & \begin{tabular}[c]{@{}c@{}}Audio \\ Generative\end{tabular} & 0 (0.0\%) & \begin{tabular}[c]{@{}c@{}}1 (3.8\%)\end{tabular} & \begin{tabular}[c]{@{}l@{}}Synthesizes audio taking \\ Mel Spectrogram(an acoustic time-frequency \\representation of a sound~\cite{melspectrogram}) as input.\end{tabular} \\ \hline

\multirow{10}{*}{\begin{tabular}[c]{@{}c@{}}Computer\\ Vision\end{tabular}} 
 & \multirow{7}{*}{Image} & Augmentation & 6 (1.6\%) & 0 (0.0\%) & \begin{tabular}[c]{@{}l@{}}Augment the images. \\ (like rotation, shearing)\end{tabular} \\ \cline{3-6} 
 &  & Classification & 94 (24.5\%) & \begin{tabular}[c]{@{}c@{}}15 (57.7\%)\end{tabular} & Classify the images according to their contents. \\ \cline{3-6} 
 &  & \begin{tabular}[c]{@{}c@{}}Feature \\ Vector\end{tabular} & 111 (29.0\%) & 0 (0.0\%) & Extract image features. \\ \cline{3-6} 
 &  & Generator & 42 (11.0\%) & \begin{tabular}[c]{@{}c@{}}2 (7.7\%)\end{tabular} & \begin{tabular}[c]{@{}l@{}}Generate images. \\ (e.g., synthesize a photo, picture style transfer,\\ enhance resolution)\end{tabular} \\ \cline{3-6} 
 &  & \begin{tabular}[c]{@{}c@{}}Object\\ Detection\end{tabular} & 4  (1.0\%) & \begin{tabular}[c]{@{}c@{}}1  (3.8\%)\end{tabular} & Find the objects in an image. \\ \cline{3-6} 
 &  & \begin{tabular}[c]{@{}c@{}}Segmentation\end{tabular} & 10  (2.6\%) & \begin{tabular}[c]{@{}c@{}}3  (11.5\%)\end{tabular} & Divide the different regions of a image. \\ \cline{3-6} 
 &  & Other & 1  (0.3\%) & 0  (0.0\%) & - \\ \cline{2-6} 
 & \multirow{3}{*}{Video} & Classification & 2  (0.5\%) & 0  (0.0\%) & Classify the videos according to their contents. \\ \cline{3-6} 
 &  & Generator & 5  (1.3\%) & 0  (0.0\%) & Generate videos. \\ \cline{3-6} 
 &  & Text & 2  (0.5\%) & 0  (0.0\%) & Extract video features. \\ \hline

\multirow{2}{*}{\begin{tabular}[c]{@{}c@{}}Natural\\ Language\\ Processing\end{tabular}} & \multirow{3}{*}{Text}  & \begin{tabular}[c]{@{}c@{}}Question\\Answering\end{tabular} & 3 (0.8\%) & 0 (0.0\%) & \begin{tabular}[c]{@{}l@{}}Answer questions in natural language\end{tabular} \\ \cline{3-6} 
 & & \begin{tabular}[c]{@{}c@{}}Embedding\end{tabular} & 99 (25.8\%) & \begin{tabular}[c]{@{}c@{}}3 (11.5\%)\end{tabular} & \begin{tabular}[c]{@{}l@{}}Changing the text (word, phrase, document) \\ into a mathematical vector.\end{tabular} \\ \cline{3-6} 
 &  & \begin{tabular}[c]{@{}c@{}}Text to\\ Mel Spectrogram\end{tabular} & 0 (0.0\%) & \begin{tabular}[c]{@{}c@{}}1 (3.8\%)\end{tabular} & \begin{tabular}[c]{@{}l@{}}Generates Mel Spectrogram \\ with natural language text\end{tabular} \\ \hline
\end{tabular}
\end{table}

Each published ML package in the studied repositories is trained with a specific algorithm on a specific dataset to help developers with a particular ML task (e.g., image classification). 
TFHub and PyTorch Hub define different type task classifications. In our research, we adopt TFHub's classification due to its clarity, and manually apply this classification on PyTorch Hub's packages. Task types unique to PyTorch Hub are added to the task type set. It should be noted that we further categorize similar task types under new created task types. 
The result of this process are the task types in Table~\ref{tab:tasktypeDistribution}, among which two are in audio processing, six are in computer vision and two are in natural language processing. 

As shown in Table~\ref{tab:tasktypeDistribution}, 
image feature vector ML models take up the largest proportion (around 29\%) in TFHub; the second and third largest task types in TFHub are text embedding (around 26\%) and image classification (around 25\%). 
In PyTorch Hub, the top three largest task type groups are text embedding (around 52\%), image classification (around 39\%) and image generator (around 3.8\%). 

Both repositories have ML models of the image classification, image generator, object detection, image segmentation and text embedding task types. TFHub has more ML models in all five types than PyTorch Hub. 
Only TFHub has ML models of audio embedding, audio pitch extraction, image augmentation, image feature vector, text question answering and video processing types. At the same time, users have to go to PyTorch Hub for ML models of audio generative (with Mel Spectrogram), text to Mel Spectrogram, and image semantic segmentation task types. 

\subsubsection{ML package Organization Practices: Family Phenomenon}
\label{sec:rq2_family}

\begin{figure} 
	\centering 
	\includegraphics[scale=0.7]{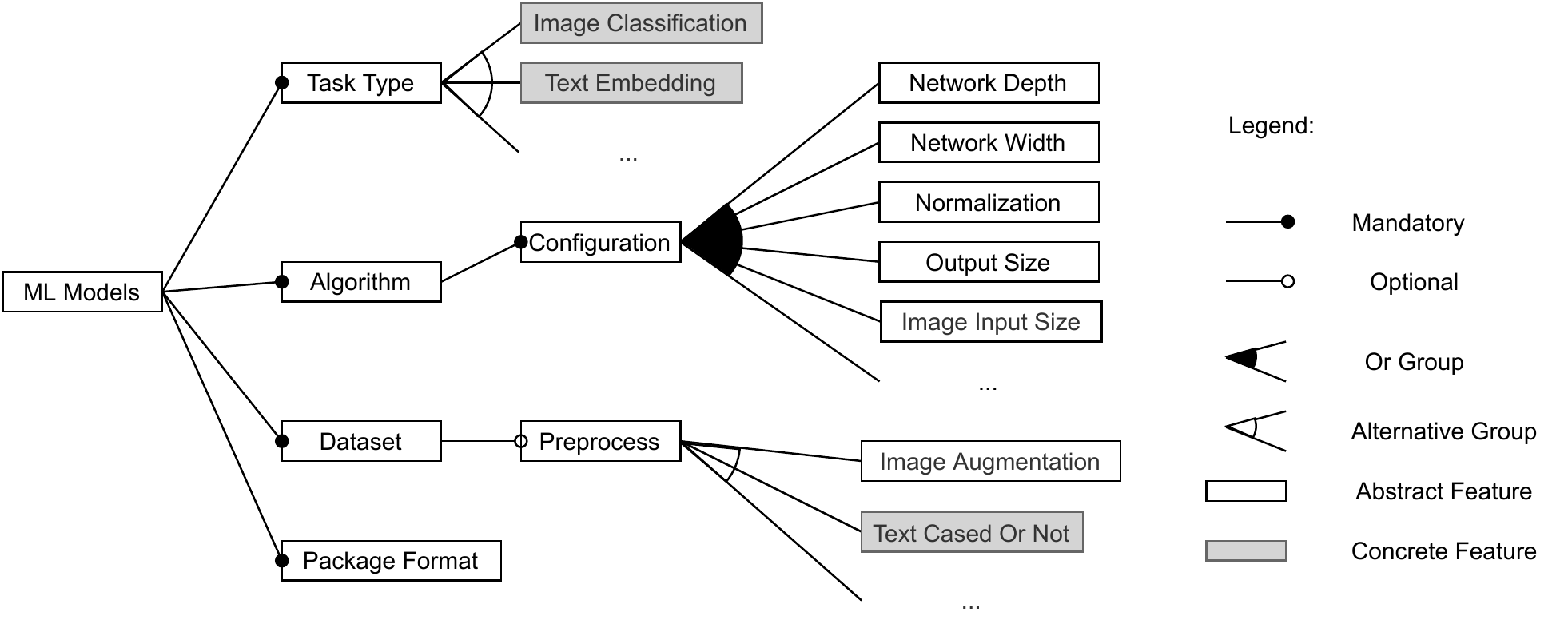} 
	\caption{The feature diagram of ML models} 
	\label{fig:feature_model} 
\end{figure}

ML packages are not organized in the same fashion in the TFHub and PyTorch Hub repositories. Although each ML package (in either TFHub or PyTorch Hub) has its individual page, we observe possible similarities among ML packages in terms of algorithm, training dataset and task type. 
Thus, in this section, we perform an in-depth analysis of the organization of these two ML package repositories through a study of the family phenomenon (introduced at the beginning of this RQ).

\paragraph{Definition}

ML packages differ from each other in terms of task types, algorithms, datasets and package formats, as illustrated in Figure~\ref{fig:feature_model}. The former three are ML-related information elements and the last one depends on the package's implementation (e.g., different frameworks provide different model formats). 

ML packages usually contain one or multiple ML models; a TFHub package always contains a single  model while a PyTorch Hub package may contain several models through the entrypoint mechanism (generally speaking, one entrypoint maps to one model). From the perspective of models, we find that some of them are the same in terms of the task type, algorithm and training dataset, but differ from each other due to different configurations of the algorithm or different pre-processing of the dataset. This phenomenon inspires us to group such models as families. 

In the context of our research, \textbf{family members have the same task type, algorithm and dataset}, but may differ in configurations, output sizes and data pre-processing. Configurations (e.g., network depth, network width, normalization, etc.) are the most common differences among family members. 
Such configurations can cause the ML models to be of different sizes in terms of FLOPs (floating point operations, a metric for the complexity of the ML model) and number of parameters; thus, having an impact on the performance and deployment of ML models. Generally speaking, the larger a ML model's size, the better its performance (like classification accuracy). However, large ML models require more computational resources and are not suitable for usage contexts like mobile phones.	
Another difference observed in family members is the output size. For image generation ML models, the output size is the size of generated images, or for text embedding ML models, the length of embedding vectors. 
Different data pre-processing steps such as text case normalization (e.g., lower and upper case and accent markers are kept or removed uniformly) are used among family members, especially in NLP ML models. 

The family phenomenon in ML package repositories is comparable to the product line architecture in traditional software engineering, which makes it easier to create closely related but varying versions of the same product~\cite{clements1999software}. It should however be noted that there is no direct mapping between model families and signatures/entrypoints. TFHub and PyTorch Hub packages are required to have signatures and entrypoints, respectively. However, their implementation mostly depends on the developer of the package. For example, a developer can create multiple entrypoints within the same package but none would use the same algorithm, dataset or task type. Thus, a package can contain models belonging to different sets of families (e.g., the \texttt{Semi-Supervised and Semi-Weakly Supervised ImageNet Models} package contains 12 models that form four families). 

Analyzing the differences between such family members provides additional understanding about ML model management and presentation; thus, it is essential that any ML package analysis considers this concept. We provide details on these different organization practices in the subsequent sections. In order to have a uniform analysis and comparison across the two ML package repositories, the subjects of the research in the subsequent subsections are rather the models within ML packages. 

\paragraph{Family Grouping Result and Analysis}

There are 43 and 28 families in the TFHub and PyTorch Hub repositories, respectively. MobileNet V1 and MobileNet V2, both trained on ImageNet, are the two largest families in TFHub with 32 and 23 members, respectively. 
TFHub model families have a median of five family members, with most families having two members. 
In PyTorch Hub, VGG Nets and BERT families have the largest number of members (eight). The median number of family members is 2.5 and most families have two members. 

\begin{figure} 
	\centering 
	\includegraphics[width=14cm]{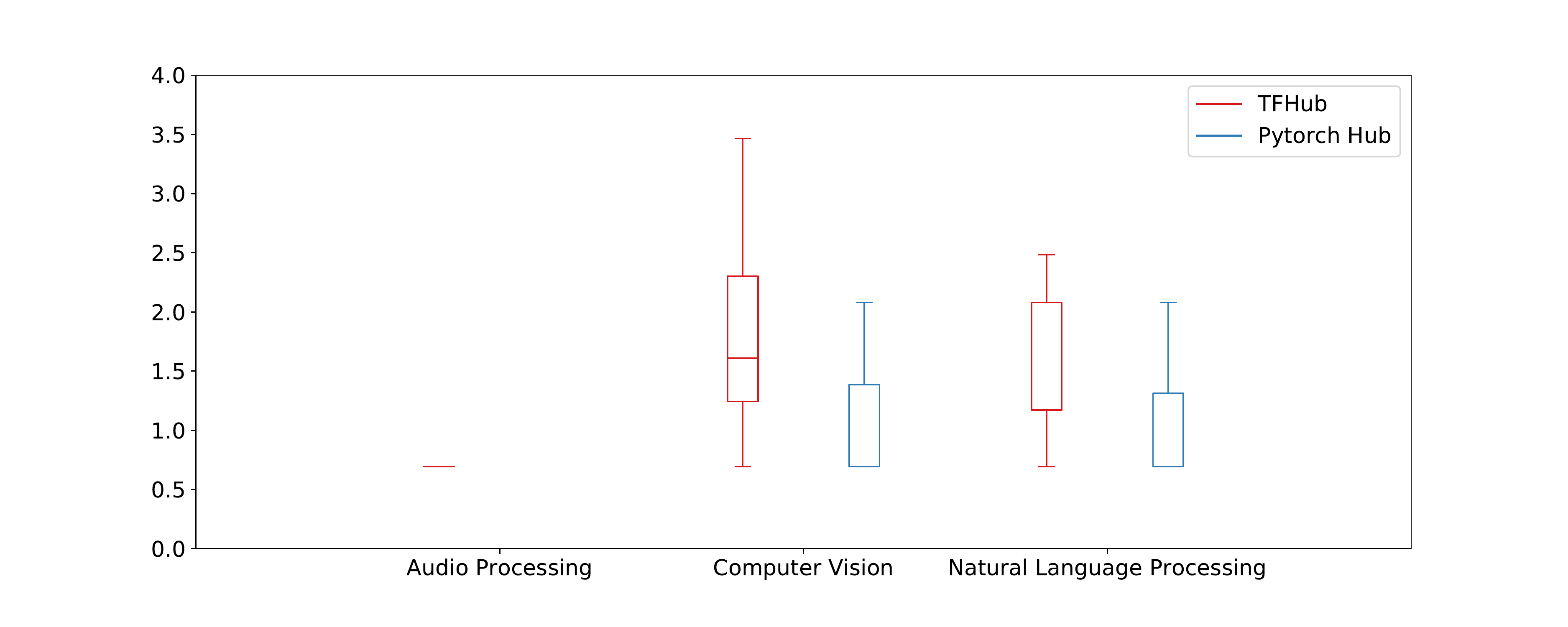} 
	\caption{Distribution of the number of family members in ML models of the studied ML package repositories based on application domain (ln-scaled). The total number of families of TFHub and PyTorch Hub are 43 and 28 respectively.} 
	\label{fig:familynumdistribution} 
\end{figure}

\begin{table}[]
\centering
\caption{Statistics of number of families and median number of family members  (only task types with at least a single family in either repository are shown)}
\label{tab:familyStatsTaskType}
\begin{tabular}{|c|c|c|c|c|c|c|}
\hline
\multirow{2}{*}{\textbf{\begin{tabular}[c]{@{}c@{}}Application\\ Domain\end{tabular}}} &
  \multirow{2}{*}{\textbf{Input}} &
  \multirow{2}{*}{\textbf{\begin{tabular}[c]{@{}c@{}}ML \\ Task Type\end{tabular}}} &
  \multicolumn{2}{c|}{\textbf{\# Families}} &
  \multicolumn{2}{c|}{\textbf{\begin{tabular}[c]{@{}c@{}}Median\\ \# Family Members\end{tabular}}} \\ \cline{4-7} 
 &
   &
   &
  \textbf{TFHub} &
  \textbf{PyTorch Hub} &
  \textbf{TFHub} &
  \textbf{PyTorch Hub} \\ \hline
\multirow{1}{*}{\begin{tabular}[c]{@{}c@{}}Audio Processing\end{tabular}} &
  \multirow{1}{*}{Audio} &
  Embedding &
  1 &
  - &
  2 &
  - \\ \cline{3-7} 
  \cline{2-7} 
  \hline
\multirow{10}{*}{\begin{tabular}[c]{@{}c@{}}Computer\\ Vision\end{tabular}} &
  \multirow{7}{*}{Image} &
  Classification &
  7 &
  13 &
  4 &
  4 \\ \cline{3-7} 
 &
   &
  \begin{tabular}[c]{@{}c@{}}Feature \\ Vector\end{tabular} &
  12 &
  - &
  2 &
  - \\ \cline{3-7} 
 &
   &
  Generator &
  7 &
  1 &
  5 &
  2 \\ \cline{3-7} 
 &
   &
  \begin{tabular}[c]{@{}c@{}}Object\\ Detection\end{tabular} &
  - &
  1 &
  - &
  2 \\ \cline{3-7} 
 &
   &
  Segmentation &
  1 &
  - &
  10 &
  - \\ \cline{3-7} 
  \cline{2-7} 
 &
  \multirow{1}{*}{Video} &
  Text &
  1 &
  - &
  2 &
  - \\ 
  \hline
\multirow{3}{*}{\begin{tabular}[c]{@{}c@{}}Natural Language\\ Processing\end{tabular}} &
  \multirow{2}{*}{Text} &
  \begin{tabular}[c]{@{}c@{}}Question\\ Answering\end{tabular} &
  1 &
  - &
  3 &
  - \\ \cline{3-7} 
 &
   &
  \begin{tabular}[c]{@{}c@{}}Embedding\end{tabular} &
  13 &
  13 &
  8 &
  2 \\ 
  \hline
\end{tabular}
\end{table}

Figure~\ref{fig:familynumdistribution} shows the distribution of the number of family members in TFHub and PyTorch Hub based on application domain. We observe the family phenomenon in only nine task types across both TFHub and PyTorch Hub (see Table~\ref{tab:familyStatsTaskType}). Six of these tasks belong to the computer vision domain (\texttt{Image Classification}, \texttt{Image Feature Vector}, \texttt{Image Generator}, \texttt{Video Text}, \texttt{Image Segmentation} and \texttt{Object Detection}), one (\texttt{Audio Embedding}) belongs to audio processing, and two (\texttt{Question Answering} and \texttt{Text Embedding}) belong to the natural language processing domain.
Audio embedding, image feature vector, image segmentation, video text and text question answering are only discovered in TFHub, while object detection family only exists in PyTorch Hub.  Some statistics about the task types are in Table~\ref{tab:familyStatsTaskType}. 

In both ML package repositories, the text embedding type has the largest number of families. TFHub has more families than PyTorch Hub in the image generator task type, while PyTorch Hub has more families of image classification task type. 


TFHub and PyTorch Hub families use 27 and 20 algorithms, respectively, with only two algorithms in common: \texttt{ResNet-V1} (image classification) and \texttt{BERT} (text embedding). On TFHub, text embedding algorithm \texttt{NNLM} has the largest number of ML models (58) while the text embedding algorithm \texttt{BERT} has the largest number of ML models (23) on PyTorch Hub. 


A great diversity of datasets is also used to train the ML models within the identified families. TFHub and PyTorch Hub ML models use 23 and 17 different datasets, respectively. Among these datasets, only three of them are common across the two repositories: \texttt{ImageNet} (image classification, image generator), \texttt{CelebA HQ} (image generation) and \texttt{Wikipedia \& BookCorpus} (text embedding). In both of the TFHub and PyTorch Hub, ImageNet is used in most of the families (20 on TFHub and 12 on PyTorch Hub). 

\paragraph{Similar Models Across ML Model Repositories}

\begin{table}[]
\centering
\caption{Similar ML models in the studied ML package repositories (* The ML model on TFHub is trained on ImageNet 2012, the PyTorch Hub model is trained on ImageNet 2014)}
\label{tab:moduleOverlap} 
\begin{tabular}{|c|c|c|c|c|c|c|}
\hline
\multirow{2}{*}{\textbf{\begin{tabular}[c]{@{}c@{}}Application\\ Domain\end{tabular}}} & \multirow{2}{*}{\textbf{Input}} & \multirow{2}{*}{\textbf{\begin{tabular}[c]{@{}c@{}}ML \\ Task Type\end{tabular}}} & \multirow{2}{*}{\textbf{Algorithm}} & \multirow{2}{*}{\textbf{Dataset}} & \multicolumn{2}{c|}{\textbf{\# Models}} \\ \cline{6-7} 
 &  &  &  &  & \textbf{TFHub} & \textbf{PyTorch Hub} \\ \hline
\multirow{4}{*}{\begin{tabular}[c]{@{}c@{}}Computer\\ Vision\end{tabular}} & \multirow{4}{*}{Image} & \multirow{4}{*}{Classification} & \begin{tabular}[c]{@{}c@{}}Inception V1\\ (GoogleNet)\end{tabular} & ImageNet* & 1 & 1 \\ \cline{4-7} 
 &  &  & Inception V3 & ImageNet & 2 & 1 \\ \cline{4-7} 
 &  &  & ResNet V1 & ImageNet & 4 & 5 \\ \cline{4-7} 
 &  &  & MobileNet V2 & ImageNet & 23 & 1 \\ \hline
\end{tabular}
\end{table}

As previously mentioned in Section~\ref{sec:rq2-tasktype}, some task types are unique to a single ML package repository. To some extent, this finding reflects the difference in terms of the contents of ML repositories. Having introduced the family concept, this section studies the similarity of the contents within the two ML package repositories. 
Table~\ref{tab:moduleOverlap} presents the number of similar ML models across the two studied ML package repositories. 

There are actually only a few ML models that overlap, see Table~\ref{tab:moduleOverlap}. For example, in TFHub, there are three ResNet V1~\cite{he2016deep} image classification ML models trained on ImageNet, their names are \texttt{imagenet/resnet\_v1\_50/classification}, \texttt{imagenet/resnet\_v1\_101/classification},  \texttt{imagenet/resnet\_v1\_152/classification} and \texttt{resnet\_50/classification}. While in PyTorch Hub package \texttt{ResNet}, there are five models \texttt{resnet18, resnet34}, \texttt{resnet50}, \texttt{resnet101}, and \texttt{resnet152} corresponding to the TFHub models.

\paragraph{Release Management of ML models} 
\label{sec:rq2_releaseManagement}

As previously discussed in RQ1, ML package repositories do not have any formalized release management or versioning mechanisms.
We also find that ML package repositories do not have a well-defined release management practice in terms of algorithm upgrades. In TFHub's ML package organization practice, a change to the algorithm leads to the creation of a new package, rather than an upgraded package. 
For example, when the algorithm used by a package is changed from \texttt{MobileNet V1}~\cite{howard2017mobilenets} to \texttt{MobileNet V2}~\cite{sandler2018mobilenetv2}, a new package page is built for the upgraded ML package. Furthermore, except for the algorithm, the two packages are the same in terms of the other two family deciding criteria (dataset and task type).
We observe several algorithm changes in TFHub such as ResNet V1~\cite{he2016deep} to V2~\cite{he2016identity}, and BERT~\cite{devlin2018bert} to ALBERT~\cite{lan2019albert}. 

Though PyTorch Hub does not support an explicit versioning mechanism, there are some cases of algorithm upgrades; an upgrade of algorithms usually leads to different entrypoints (different models) in the same package rather than different versions. Examples of observed algorithm upgrades are BERT to distillBERT~\cite{sanh2019distilbert} and RoBERTa~\cite{liu2019roberta}, and GPT~\cite{radford2018improving} to GPT-2~\cite{radford2019language}.

\subsection{Implications}


\noindent
\textbf{Release Management}. 
Currently, the upgrade of ML package's algorithm is not a versioned change. 
It is worthwhile for ML package repositories to consider new versions of an algorithm as an upgrade. There are two benefits: 
(1) Users are better aware of how many versions of an algorithm to choose from. It is a good practice to provide ML package users with information transparency. For example, without our research, TFHub users may not easily know that there are two versions of MobileNet algorithm, two versions of ResNet algorithm and three versions of Inception algorithm. If the number of ML packages keeps growing, this transparency will be more helpful for users. 
(2) This practice helps users better understand a group of algorithms and help them narrow down the search scope. For example, although all MobileNet ML packages are suitable for mobile platform deployment, MobileNet V2 being better than V1 guides users to choose ML packages directly from V2 ones, rather than trying out from V1 ones. 

Compared to traditional software package evolution, the evolution of an ML package can involve any of these things: (1) algorithm update, (2) update to non-algorithm related code (e.g., command line arguments, tuning, etc.), (3) changes to data, and (4) changes to input/output tensors. Among them, (1) and (3) are ML package specific, whereas (2) is common for all software projects. (4) can be applicable under the model family phenomenon. Furthermore, some of these are orthogonal for ML packages. For example, one product may change the algorithm but keep the data as it is, or update both the algorithm and the data. Some products may have multiple variants trained on different data. The aforementioned issues pose several research opportunities in the ML domain. Thus, there is the need for ML researchers to identify existing release management approaches in the traditional software engineering and product line domains that can be adopted and extended.\\

\noindent
\textbf{Impedance mismatch between model families and software package distribution/versioning}. There are many families in the same task type, and there are many members in a family. Though this provides users with a great diversity of ML models, the similarity between families and members makes choosing right ML model be difficult and confusing for users without solid ML expertise (like general software engineer and non-ML researchers). In the worst case, users may need to try the different model families, and probably each ML model within a family, to identify trade-offs between performance and computational resource consumption. 


Based on our findings, we observe that the unit of shipping models is not that straightforward and formalized: should a package contain a whole family or a subset of members (based on entrypoints), or even multiple families?
In addition, the family phenomenon may introduce some other software engineering challenges. For example, how are models in a family upgraded? How are changes within these families managed as the models evolve over time? 
Given the inherent similarities with the family phenomenon, ML researchers should seek to adopt the advanced practices of the traditional software product line architecture domain~\cite{botterweck2014evolution}. 
\section{RQ3: What is the process needed in order to use the functionalities from software/ML package repositories?}
\label{sec:rq3}

Having identified the information within the studied ML package repositories (RQ1), as well as their organization practices (RQ2), this RQ investigates the processes needed to reuse these shared ML packages. First, we examine the basic functionalities and the package usage practices supported by the ML package managers (e.g., \texttt{tensorflow\_hub} library~\cite{tensorflowhublibrary} for TFHub, PyTorch Hub API in PyTorch library). The functionalities and practices will be compared against those of software package managers, whenever applicable, to understand the commonalities and uniqueness between them. 
Next, we study the different usage contexts supported by ML libraries from their documentation. 

\subsection{Approach}
\label{sec:rq3-approach}

Package managers provide some basic functionalities such as installing, upgrading and removing packages within a programming language-specific development environment~\cite{di2008package}. Due to the longer existence and wider usage of software package repositories, we regard the software package managers and their APIs as a baseline, and we attempt to identify such similar functionalities for ML libraries. 
	
First, we look for documentation and online tutorials about the three functionalities (installation, upgrade, removal) of both ML package repositories and software package repositories. If no corresponding materials of a functionality are found for a given repository, we regard this functionality to be unsupported. Table~\ref{tab:basicFuncUseInfo} summarizes the basic functionalities, the supported package formats and primary usage mechanism of the studied package repositories. 
Next, we also identify the different supported usage contexts and the steps needed to use the packages of the ML and software package repositories.

\subsection{Findings}

\subsubsection{Basic functionalities  of ML and Software package managers (Installation, Upgrade, Removal)}
\label{sec:rq3-basicfunction}



\begin{table}[]
\centering
\caption{Basic functionalities and usage information of software and ML package repositories}
\label{tab:basicFuncUseInfo}
\begin{tabular}{|c|c|c|c|c|c|c|}
\hline
\multicolumn{2}{|c|}{\multirow{2}{*}{\textbf{Repositories}}} & \multicolumn{3}{c|}{\textbf{Package Manage Functionalities}} & \multirow{2}{*}{\textbf{\begin{tabular}[c]{@{}c@{}}Supported Format(s)\\of Packages\end{tabular}}} & \multirow{2}{*}{\textbf{\begin{tabular}[c]{@{}c@{}}Usage\\Mechanism\end{tabular}}} \\ \cline{3-5}
\multicolumn{2}{|c|}{} & \textbf{Install} & \textbf{Upgrade} & \textbf{Remove} &  &  \\ \hline
\multirow{3}{*}{Software} 
 & npm & \cmark & \cmark & \cmark & \begin{tabular}[c]{@{}c@{}}Various\\ (code file)\end{tabular} & API \\ \cline{2-7} 
 & PyPI & \cmark & \cmark & \cmark & \begin{tabular}[c]{@{}c@{}}Standard \\ (.tar.gz,  .whl)\end{tabular} & API \\ \cline{2-7} 
 & CRAN & \cmark & \cmark & \cmark & \begin{tabular}[c]{@{}c@{}}Standard \\ (zipped package)\end{tabular} & API \\ \hline

\multirow{2}{*}{ML} 
 & TFHub & \cmark &  &  & \begin{tabular}[c]{@{}c@{}}Various\\ (hub.Module, \\ TF2 SavedModel,\\ other formats)\end{tabular} & Signature \\ \cline{2-7} 
 & PyTorch Hub & \cmark &  &  & \begin{tabular}[c]{@{}c@{}}Mostly Standard\\ (GitHub repository + \\ .pt/.pth/unknown format)\end{tabular} & Entrypoint \\ \hline 
\end{tabular}
\end{table}


Though users have to follow similar steps before using packages within software and ML package repositories, there are, however, a few significant differences in how these steps are implemented for software and ML packages. It generally takes 2 steps to use any software or ML package: (1) installing/upgrading packages, and (2) invocation of the functionalities from a loaded/imported package.
	
\begin{figure}[htbp]
\centering                                                          
	\subfigure[PyPI]{
		\begin{minipage}{7cm}
			\centering                                                          
			\includegraphics[scale=0.7]{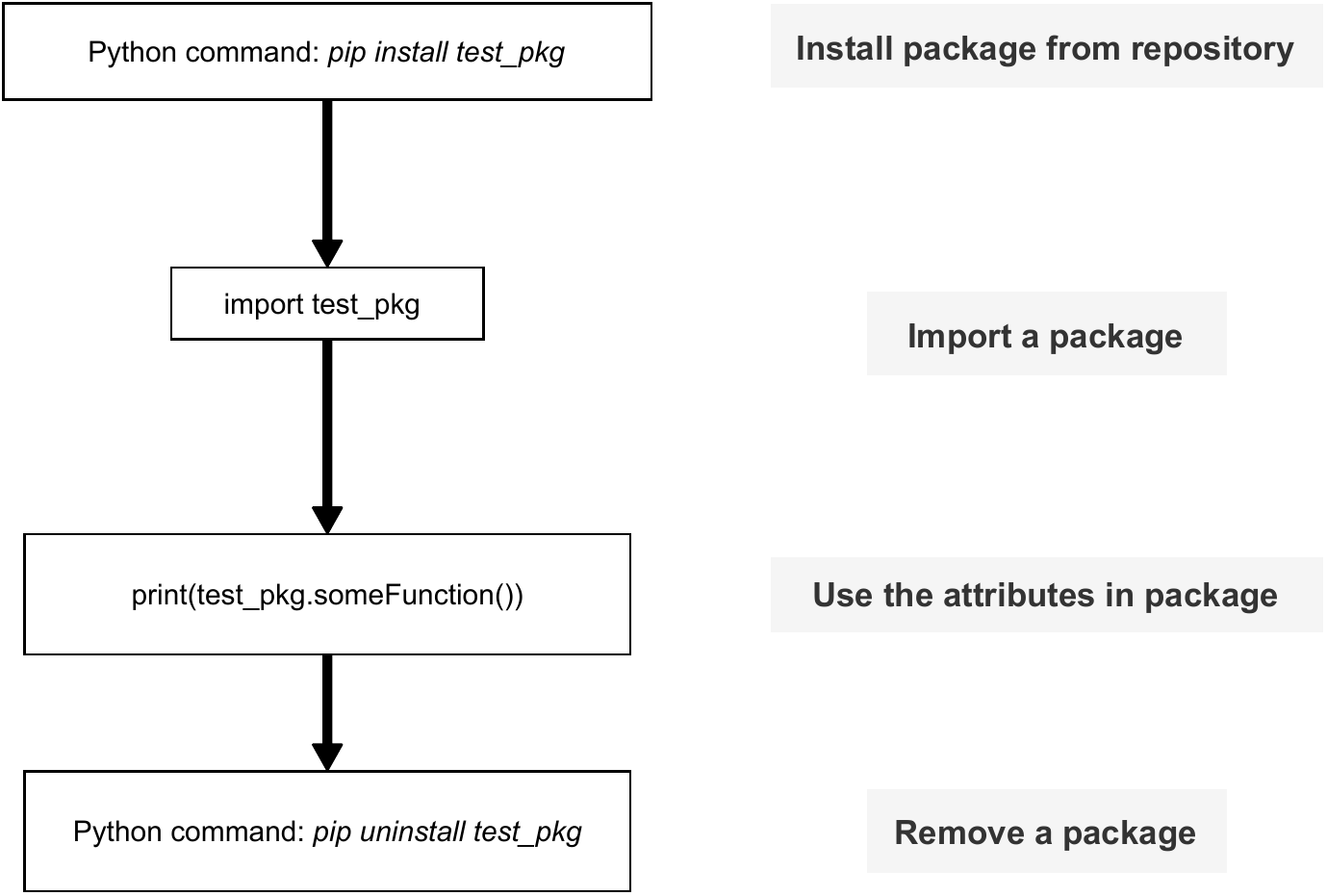} 
			\label{fig:npmPyPICRANLoadUse}              
		\end{minipage}
	}
	\subfigure[TFHub]{
		\begin{minipage}{7cm}
			\centering                                                          
			\includegraphics[scale=0.7]{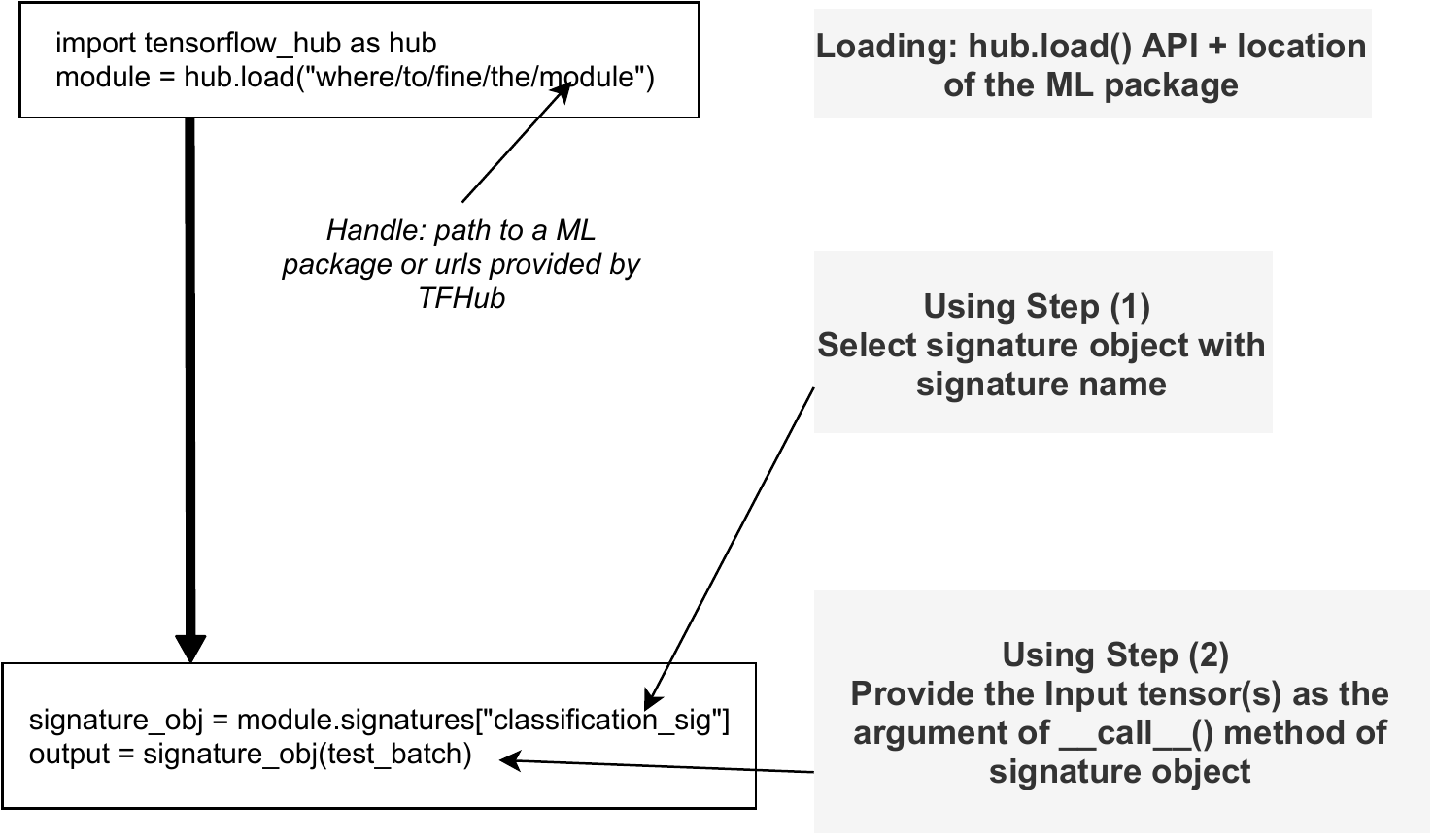}     
			\label{fig:hubModule-LoadUse1Exp}           
		\end{minipage}
	}
	\subfigure[PyTorch Hub]{
		\begin{minipage}{7cm}
			\centering                                                          
			\includegraphics[scale=0.7]{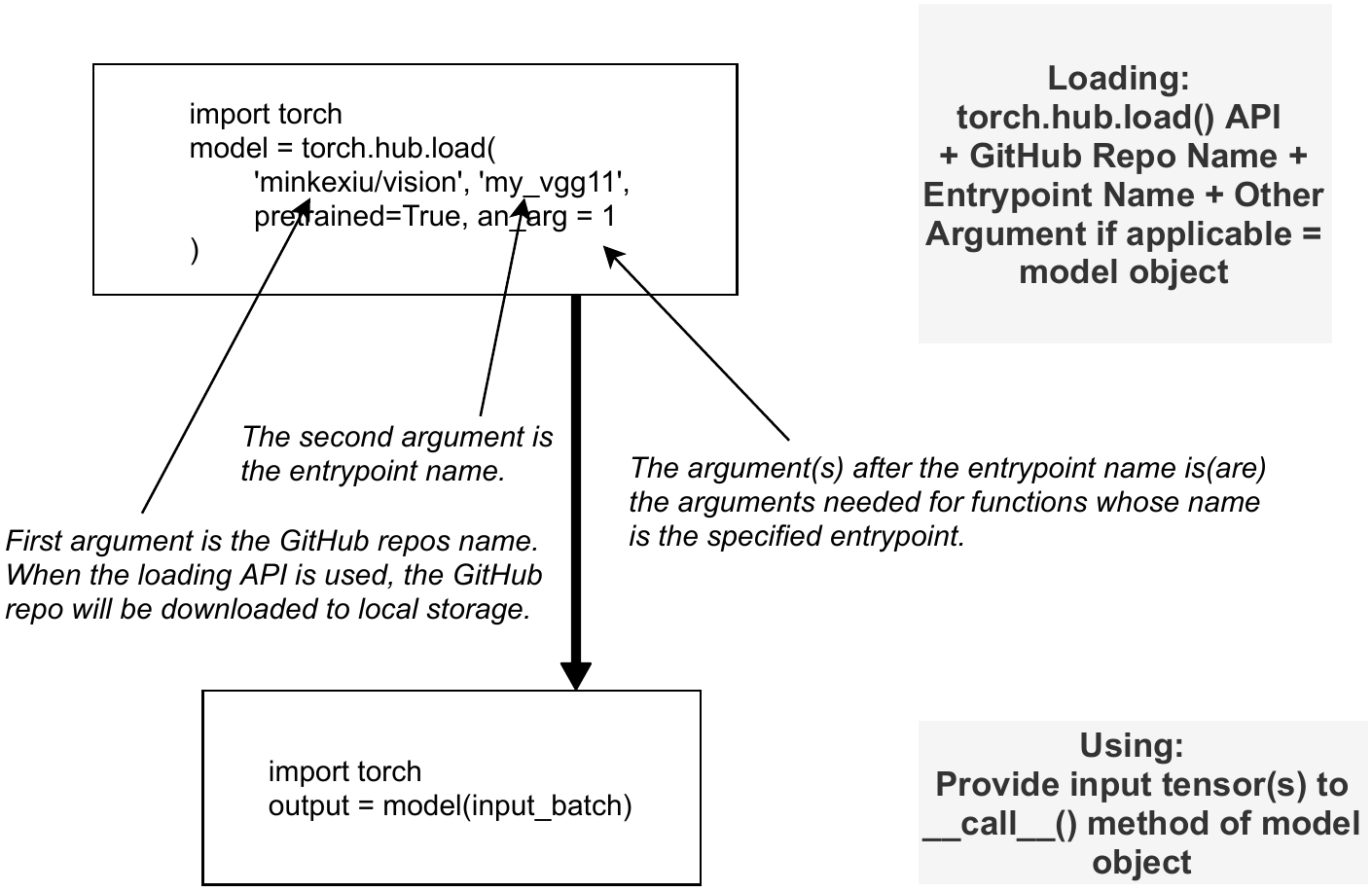}    
			\label{fig:PyTorchHubProduceLoadUse-LoadUse}             
		\end{minipage}
	}
\caption{The process of loading and using packages} 
\label{fig:1}                                                        
\end{figure}

Software packages are mostly installed via terminal commands provided by their package managers (see Figure \ref{fig:npmPyPICRANLoadUse}). However, there is no clear division between installation and loading in ML package repositories. ML packages require a runtime load step that downloads the ML artifacts if not yet in cache, and selects the right model for further use. 
For example, figure~\ref{fig:hubModule-LoadUse1Exp} demonstrates how to load a ML package from TFHub. 
As shown in the figure, TensorFlow's \texttt{tensorflow\_hub} library, which is the library that mainly supports the ML package management functionalities and ML package usage, provides an initialization API called \texttt{hub.load()}. 
This API takes in an argument for the location of the ML package; this can be either a link to TFHub pages, a link to some specified online zip file or a path to local package. Given the location, the library downloads the package (if not already in the cache), loads it, and makes it ready for use. A similar practice is used to load PyTorch Hub packages, as shown in  Figure~\ref{fig:PyTorchHubProduceLoadUse-LoadUse}. 

Additionally, there is no real support for package upgrades in ML package repositories. Unlike software packagers that provide users with package upgrade commands, users of ML packages have to manually specify the version of a package to load at runtime. This practice is further exacerbated due to the lack of formal package versioning mechanisms. For example, users of TFHub packages need to specify the version number as part of the url (e.g., the ``\texttt{../4}'' at the end of \texttt{https://tfhub.dev/google/imagenet/mobilenet\_v1\_050\_160/classification/4} shows the version). Given that PyTorch Hub has no versioning support for packages, users must decide at runtime whether to download an updated PyTorch package from its GitHub repository$\footnote{The source code of PyTorch Hub packages are stored on GitHub. The links to model files (\texttt{.pt}, \texttt{.pth} files) will be stored in the GitHub code but the model files themselves may be stored on some other places rather than GitHub.}$ or use an existing cached copy. 
As a result, a different GitHub snapshot of a model can be loaded each time, introducing severe inconsistency problems to users. 

Also, we could not find any similar functionality for the removal of packages provided by ML libraries. The package initialization API of PyTorch has an argument that if set, would always download a new package even if there is a cached one. But this is not a real removal functionality as users need to manually remove the cached models eventually.

\subsubsection{Package Usage}

Despite the close similarities of how ML and software packages are installed or loaded, the process of actually using these packages is different. Once software packages are installed, the users only need to import the package and utilize the various APIs within their system. ML packages, on the other hand, require an additional step after they are loaded before they can be used. This difference is brought by the signature mechanism of TFHub packages and entrypoint mechanism of PyTorch Hub packages. 
An in-depth discussion of how ML packages are used under these two mechanisms is provided below. 

\paragraph{TFHub Signature Mechanism} 
\label{sec:rq3-aihubloadanduse}


A signature is a particular combination of input and output data-structures (also called tensors
~\cite{defOfTensor}) used by a ML model. Given that some ML packages can be used for more than one task, the signature mechanism is used to allow users to express the task to perform. For example, \texttt{imagenet/mobilenet\_v1\_050\_160/classification$\footnote{https://tfhub.dev/google/imagenet/mobilenet\_v1\_050\_160/classification/4}$} is a package which allows users to perform either image classification or feature extraction in a set of images. 
Thus, a user performing an image classification task with the \texttt{imagenet/mobilenet\_v1\_050\_160/classification} package must provide the input tensor in the expected shape before the correct output tensor will be returned. 

\paragraph{PyTorch Hub Entrypoint Mechanism} 
\label{sec:rq3_pthubpackage}

When loading a PyTorch package for use, the \texttt{torch.hub.load()} API requires, in addition to the GitHub repository containing the necessary code of this package, the name of an entrypoint name and any additional arguments needed by the provided entrypoint. 
An entrypoint is essentially a Python function defined in the package's source code that implements a particular configuration of an algorithm. It should be noted that how the function actually works is totally decided by developers, including the argument settings, implementation logic, and where to find the pre-trained model files. 

\begin{figure} 
	\centering 
	\includegraphics[scale=0.5]{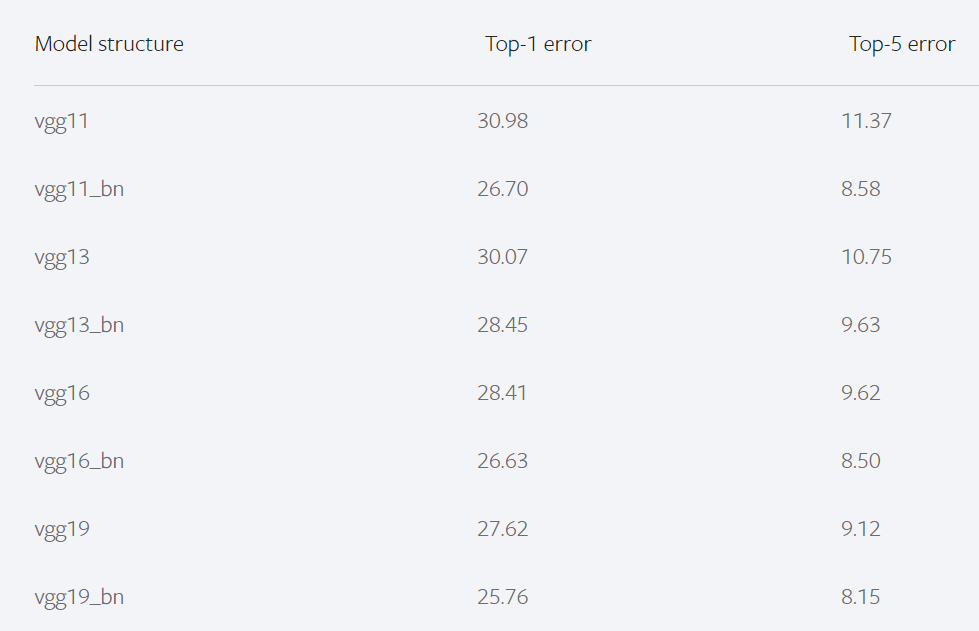} 
	\caption{An example of the list of entrypoints of a PyTorch Hub package with their respective quality measures} 
	\label{fig:entrypointlist} 
\end{figure}

We observe that currently the PyTorch Hub entrypoint mechanism is not as formalized as the TFHub signature mechanism. 
Several packages on PyTorch Hub do not provide a complete entrypoint list (an example of how entrypoints are displayed for a package is shown in Figure~\ref{fig:entrypointlist}); one has to manually search within the source code repository$\footnote{https://pytorch.org/hub/pytorch\_fairseq\_translation/}$. 
Secondly, 
package developers may implement the entrypoint differently given the lack of formalism or best practices. Such differences in entrypoint definitions makes the loading process of a package difficult for users. 
The aforementioned two issues show the need for the package developers to prepare sufficient documentation that explains the full functionality and loading process of their packages, rather than expecting general users to read and understand the source code. In comparison, TFHub packages do a better job by providing APIs to inform the users of the signatures within packages.

\paragraph{Signature/Entrypoint mechanism vs. Model families}

We observe that the variation within ML packages can be viewed along different dimensions: model families (see RQ2) and signatures/entrypoints. There may be a number of models (in a the same family) that use the same algorithms and datasets but different training hyper-parameter configurations or computational features (e.g., with or without batch-normalization hyper-parameters). We could also have different variations of an ML package defined through entrypoints.

However, we observe that although the number of entrypoints in a PyTorch Hub package indicates the number of different  models in the package, not all entrypoints in a package belong to the same family; a PyTorch Hub package can have multiple families, each consisting of a subset of its entrypoints.
In TFHub, each package consists of exactly one model which can have multiple signatures, another form of ML-specific variation. Thus, there seems to be no relation between signatures and model families. 

Consequently, the development process of ML packages can be considered to be analogous to the concept of product line architectures. 
Such product line architecture-like practices of ML packages bring some benefits. First, it makes the organization of different variants easier. For example, without the signature mechanism, the multiple signatures have to be independent packages, causing redundancy and the synchronization of their changing and maintenance will take a lot of efforts. Secondly, such mechanism help the users to easily compare the functionalities of related packages.  

\begin{table}[]
\centering
\caption{TensorFlow and PyTorch package/model usage contexts}
\label{tab:tf-pt-deployment}
\begin{tabular}{|c|c|c|c|c|}
\hline

\multirow{2}{*}{\textbf{\begin{tabular}[c]{@{}c@{}}Deployment\\ Target\end{tabular}}} & 
\multicolumn{2}{c|}{\textbf{\begin{tabular}[c]{@{}c@{}}Typical Model\\ Format\end{tabular}}} & \multicolumn{2}{c|}{\textbf{\begin{tabular}[c]{@{}c@{}}Call (Serve)\\ Locally or Remotely\end{tabular}}} \\ \cline{2-5} 
 & \textbf{TensorFlow} & \textbf{PyTorch} & \textbf{TensorFlow} & \textbf{PyTorch} \\ \hline

Python & \begin{tabular}[c]{@{}c@{}}Checkpoint, \\ SavedModel, \\ Frozen GraphDef, \\ hub.Module\end{tabular} & (All the formats) & Locally & Locally \\ \hline
Edge Device & \begin{tabular}[c]{@{}c@{}}FlatBuffers \\ (iOS, Android, RPi)\\ (Swift, Objective-C)\end{tabular} & \begin{tabular}[c]{@{}c@{}}.pt File\\ (iOS, Android)\end{tabular} & Locally & Locally \\ \hline
JavaScript & \begin{tabular}[c]{@{}c@{}}TensorFlow.js Model\\ (Browsers, node.js)\end{tabular} & - & Locally & - \\ \hline
\begin{tabular}[c]{@{}c@{}}Remote\\ Device\end{tabular} & SavedModel & .pt, .pth File & Remotely & \begin{tabular}[c]{@{}c@{}}Remotely\\ (Flask\\ Needed)\end{tabular} \\ \hline
\begin{tabular}[c]{@{}c@{}}Other \\ Languages\end{tabular} & \begin{tabular}[c]{@{}c@{}}SavedModel,\\ .pb File\\ (C++, Java, Go, etc.)\end{tabular} & \begin{tabular}[c]{@{}c@{}}TorchScript\\ (C++, Java)\end{tabular} & Locally & Locally \\ \hline
\begin{tabular}[c]{@{}c@{}}Other ML\\ Frameworks\end{tabular} & - & \begin{tabular}[c]{@{}c@{}}ONNX\\(Caffe2, MXNet, \\CNTK, etc.)\end{tabular} & - & Locally \\ \hline
\end{tabular}
\end{table}

\begin{table}[]
\centering
\caption{Advantages and disadvantages of different formats of packages/models}
\label{tab:prosandcons}
\begin{tabular}{|c|c|c|c|}
\hline
\textbf{ML Framework} & \textbf{\begin{tabular}[c]{@{}c@{}}Package/\\Model Format\end{tabular}} & \textbf{Advantages} & \textbf{Disdvantages} \\ \hline
\multirow{6}{*}{TensorFlow} & checkpoint & Suitable for recording training process. & Not suitable for deployment. \\ \cline{2-4} 
 & SavedModel & \begin{tabular}[c]{@{}c@{}}Standard model saving format. \\Suitable for deployment.\end{tabular} & May not support specialized usage context. \\ \cline{2-4} 
 & Frozen GraphDef & \begin{tabular}[c]{@{}c@{}}Size saving. \\ Suitable for inference-only usage.\end{tabular} & Parameters in it cannot be changed. \\ \cline{2-4} 
 & hub.Module & Specified for model sharing on TFHub. & Out of date in TensorFlow 2.x era. \\ \cline{2-4} 
 & TensorFlow Lite Model & Can be optimized in size. & Optimization may reduce performance. \\ \cline{2-4} 
 & TensorFlow.js Model & Can be used in JavaScript environment. & \begin{tabular}[c]{@{}c@{}}May have less support\\than traditional TensorFlow.\end{tabular} \\ \hline
\multirow{3}{*}{PyTorch} & .pt, .pth file & Standard model saving format. & May not support specialized usage context. \\ \cline{2-4} 
 & TorchScript & Can be used in multiple languages. & \begin{tabular}[c]{@{}c@{}}Inappropriate for internal\\model deployments.\end{tabular} \\ \cline{2-4} 
 & ONNX & Can be used by different ML frameworks. & May not support specialized usage context. \\ \hline
\end{tabular}
\end{table}

\subsubsection{Usage Contexts Supported by ML Libraries}
\label{rq3-deployment}


TensorFlow and PyTorch provide support for using packages in different contexts. Usage contexts are the specific deployment platforms or software environments upon which ML packages are loaded and called. Each usage context prioritizes certain attributes of an ML package such as size, performance and portability. 
As such, ML packages and models used in different contexts are usually in different formats. It should be noted that in this subsection, our analysis is focused at the level of the models within the ML packages, rather than the packages themselves. 
Table~\ref{tab:tf-pt-deployment} shows the various usage contexts available to TensorFlow and PyTorch packages. 
In the table, each row represents a group of relevant deployment scenarios. 
The first column shows the usage contexts supported by TensorFlow and PyTorch. It should be noted that \texttt{edge device} and \texttt{remote device} have some overlap, e.g., TensorFlow models may be run on Python environments on both Raspberry Pi (as a kind of edge device) and web servers (as a kind of remote device). The reason for splitting these two rows is to emphasize their most representative characteristics, i.e., computational resource limitation for edge devices and different calling mechanisms for remote server. The second and third columns represent the typically used model formats within the given deployment environment (based on the ML package repository). The fourth and fifth column explain how the models are called (or served). 
If the loading and the whole usage process happen on the same machine, it is considered ``local''.  
However, if the package is loaded on another machine and the usage process need remote communication between different machine, it is considered ``remote''. For example, a user can use HTTP requests to obtain inference results from an image classification package deployed on a remote server. 
Both ML frameworks have five deployment scenario groups. In general, TensorFlow supports more usage contexts, which can be attributed to TensorFlow's longer existence. 

Unsurprisingly, TensorFlow and PyTorch have the best support for integration directly into a Python code base. 
It is a general practice that a model is developed on Python and deployed in other scenarios. 
For TensorFlow, there are four common saved model formats: checkpoint$\footnote{https://www.tensorflow.org/guide/checkpoint}$, SavedModel$\footnote{https://www.tensorflow.org/guide/saved\_model}$, Frozen GraphDef$\footnote{https://github.com/tensorflow/tensorflow/blob/master/tensorflow/python/tools/freeze\_graph.py}$ and hub.Module$\footnote{https://www.tensorflow.org/hub/api\_docs/python/hub/Module}$. 
Checkpoint is suitable for temporarily save the training process, generally only contains parameter values but not calculations. So this character makes it not very suitable for sharing because the code of algorithm should be separately provided. 
SavedModel can save both parameters and calculations, it can be used off the shelf without any algorithm code. This character makes it suitable for deployment and sharing. In TensorFlow 2.x it is also the general format for saved model. 
Frozen GraphDef is extremely lightweight and suitable for deployment and doing inference. But its parameter values cannot be changed so models in this format cannot be further trained or fine-tuned. 
hub.Module is specially invented format for model sharing on TFHub and it's being replaced in TensorFlow 2.x API era. 
As for PyTorch, the common saved model formats under Python environment are \texttt{.pt} and \texttt{.pth}. But shared models in PyTorch Hub package may not be very suitable for deployment because the parameter values (in \texttt{.pt} or \texttt{.pth} files) and algorithms (in code base) are separately stored. A summary of the advantages and disadvantages of each model format is provided in Table~\ref{tab:prosandcons}.

TensorFlow and PyTorch models also can be deployed on edge devices, like mobile platforms, 
which suffer from limited computational and storage resources. TensorFlow provides a set of tools in TensorFlow Lite\cite{tflite} for deployment on edge device. Normal TensorFlow Python models can be converted into a TensorFlow Lite format model named FlatBuffers$\footnote{https://www.tensorflow.org/lite/convert/index}$. In addition, both TensorFlow and PyTorch provide some quantization methods, like saving the values in lower precision, that make models smaller and minimize the degradation of performance. 

TensorFlow and PyTorch models can be deployed for use in other languages. 
For TensorFlow, the SavedModel can be loaded by the TensorFlow API in C++, Java, Go, Swift, etc. PyTorch only supports C++ and Java in its documentation currently. PyTorch uses \texttt{TorchScript} to represent a model that is independent from Python and can be loaded and executed by PyTorch API in C++.  

TensorFlow and PyTorch models can be served remotely and called through a REST API for inference. For TensorFlow, a served model will be in SavedModel format. The containerized serving process for TensorFlow models is introduced by TensorFlow's documentation in detail. In addition to REST API calling, served TensorFlow models can also be called through gRPC~\cite{grpc}, a high performance RPC framework. For PyTorch, the models needs to be deployed and served with the help of Flask\cite{flask}, a third-party lightweight web framework in Python. 

TensorFlow and PyTorch each has some unique usage contexts as well. TensorFlow models can be deployed on JavaScript environments like browsers and Node.js. TensorFlow in JavaScript is called TensorFlow.js\cite{tfjs}. The TensorFlow.js models can be converted from normal Python-based models. PyTorch models can be exported into ONNX format\cite{onnx} and loaded by any ONNX-supporting frameworks, like Caffe2, MXNet, CNTK etc.. 

From the Table~\ref{tab:tf-pt-deployment}, we found that the usage contexts are more complex in ML packages than in software packages. Software packages can be deployed successfully within any usage context supports the languages and they do not have much variation in formats. For example, a Python package can be deployed on any platforms or environments that support Python, and the Python package will always be in \texttt{.whl} or \texttt{.tar.gz} formats$\footnote{https://packaging.python.org/tutorials/packaging-projects/\#generating-distribution-archives}$. 
However, this is not the case for ML packages/models. When a user wants to use a model in different contexts, the model formats will vary and the variation may lead to different loading and usage practices. For example, different APIs are needed to load and use TensorFlow.js and SavedModel model formats, or a SavedModel may need to be converted to a TensorFlow.js format before it is used in a browser. Though package/model formats can be converted between each other, considering the most suitable format in advance can at least save the conversion effort.

\subsection{Implications}



\textbf{Usage Context}. 
Unlike software packages that have similar usage steps, irrespective of the OS or hardware of a system, this is not the case for ML packages. There are many usage contexts for ML packages and models. 
The considerations for ML packages in different usage contexts are more than those for software packages. The loading and usage processes may differ according to the usage context (e.g., different sets of APIs, different formats of saved models). This characteristic is another reason for the users (like software engineers and data scientists) to consider the usage context earlier and make suitable trade-offs.\\

\noindent 
\textbf{Pre-defined interfaces for ML packages (signatures and entrypoints)}. 
ML libraries use special mechanisms for loading and using a package: \texttt{signatures} for TFHub packages and \texttt{entrypoints} for PyTorch Hub packages. These mechanisms are different from software packages that can be simply imported and then used. 
Those mechanisms adopt the product line architecture from traditional software engineering and leverage some of its benefits. 
Though these mechanisms provide the needed flexibility for creating and using variants of an ML package, they also introduce a steep learning curve for new ML developers, especially traditional software engineers.\\


\noindent
\textbf{Package Management Functionalities}. These two aspect of the ML package repositories are still a work in progress.
Currently, ML package repositories only support package installation. The upgrade, and removal functionalities are either missing or have limited support. In the future, TensorFlow and PyTorch maintainers can think about developing the actual ML package upgrade and removal functionalities.\\

\noindent
\textbf{Package Documentation}. There is limited documentation on how users (especially with limited ML experience) can integrate pre-trained models into their applications.  
Given the relative complexity of ML packages, in comparison to software packages, the documentation of ML packages are expected to be very detailed. Loading a package requires a lot of information in terms of versions, signatures and entrypoints. Such details need to be in well-described. For example, the documentation of PyTorch Hub packages' entrypoints are not complete (5 out of the 15 packages with multiple entrypoints do not provide a full list of their entrypoints). 
Contributors of the PyTorch Hub and TFHub packages need to treat the documentation highly to serve the users better. Researchers can also propose code summary techniques for generating detailed and formalized documentation, and tools to ensure that strict naming conventions for parameters can are adhered to.\\ 
\section{Related Work}
\label{sec:relatedWorks}

Here we discuss two areas of prior work related to this paper: empirical studies on software package repositories and sharing of reusable ML packages. 

\subsection{Software Package Repositories}
\label{sec:relatedwork-pkgRepoStudies}

\textbf{Exploratory Study}: There are a few exploratory studies on software package repositories. Bommarito et al.~\cite{bommarito2019empirical} analyzed the basic data of all of the packages on PyPI at that time, including information elements like packages, releases, dependencies, category classifications, licenses, package imports, authors, maintainers, and organizations. They reported the evolution of the PyPI repository in terms of active packages, new authors and new import statements. They observed highly right-skewed distributions of package release numbers, authors' package and release numbers, package import numbers, size of packages and releases. They also found that most of the packages are contributed by single individuals. Raemaekers et. al.~\cite{raemaekers2013maven} presented a dataset that contains code metrics, dependencies, breaking changes between library versions of more than 148 thousand jar files and a complete call graph of the entire Maven repository. \\

\noindent
\textbf{Dependency Management}: Dependency management on software package repositories is an aspect that attracted lots of prior work. 
Some researchers have studied the impact of dependencies (from software package repositories) on project health. Alqahtani et. al.~\cite{alqahtani2016tracing} used a unified ontological representation to establish bi-directional traceability links between security vulnerability databases and software software repositories. It is shown that when packages are shared, knowledge, information and vulnerabilities are also shared. Eghan et. al.~\cite{eghan2019api} took Maven as research subject and found that dependencies on external libraries have an impact on project quality in terms of security vulnerabilities, license violations, and breaking changes. 

Decan et. al.~\cite{decan2016github} conducted a study about R packages distributed on CRAN and GitHub and found that on GitHub, which is an increasingly used R package distribution platform, packages are subject to inter-repository dependency problems that interfere their automatic installation. Cogo et al.~\cite{cogo2019empirical} looked at the phenomenon that developers downgrade the dependencies in the npm repository. They found the reasons behind the occurrence of downgrades, how the versioning of dependencies changed when downgrades occur and how fast downgrade occurs. 

Valiev et al.~\cite{valiev2018ecosystem} explained that the interdependent network of open source projects is the software repository, the sustainability (maintainability, attractive to new comers, economic value of the project, etc.) of the projects that comprising an repository may be determined by the repository context as well. Through the case study of the PyPI repository, they found that project ties and relative position in dependency network have impact on sustained project activity.  Abdalkareem et al.~\cite{abdalkareem2020impact} calculated the proportion of trivial packages in npm and PyPI package repositories. They surveyed the developers about the reasons and drawbacks of using trivial packages. They also found that only part of trivial packages are tested, and few studied trivial packages have more than 20 dependencies. \\

\noindent
\textbf{Quality}: In addition to dependency management, prior work also studied the quality aspects of various software package repositories. For example, Claes et al.~\cite{claes2015empirical} studied the phenomenon that developers copy the code from CRAN packages to their code rather than depend on packages. They learned the characteristics of the evolution of cloned code and the reasons behind the cloning activity. 
Trockman et al.~\cite{trockman2018adding} explained that project maintainers use badges to signal the quality of their projects to contributors and users on social coding platforms. Their investigation into the badges in npm repository identifed the key quality attributes of interest to project maintainers and how well these qualitity attributes are reflected by badges. 

\subsection{Sharing Reusable ML Packages}

Currently there is limited research in this area. Research mentioning TFHub and PyTorch Hub are mostly using the ML packages from those hubs or contributing new ML packages to them. 
Touvron et al.~\cite{touvron2019fixing} proposed an image classification optimization strategy relying on the fine-tuning of PyTorch Hub pre-trained ML packages. 
Yang et al.~\cite{yang2019multilingual} proposed a new NLP algorithm and published the pre-trained ML packages on TFHub. 
Braiek et. al.~\cite{BraiekMSR2018} examined the evolution of ML frameworks and their repository (actors and adoption over time) to understand the role of open-source development in modern ML. They found that ML is between the early adoption and early maturity stage. They also found that companies are the main drivers of open-source ML, with the development teams consisting mostly of engineers and industry scientists. They also identify that the big cloud computing companies introduce a risk of a vendor lock-in for future ML development.  

While our previous work~\cite{xiu2020exploratory} studied the structure and the contents of ML model stores, in which users have to pay for the cloud-based ML model distribution, this work is the first exploratory study focusing on the functionalities, information and offerings on the ML package repositories on which packages are freely distributed through ML package manager (related ML libraries). 
The combination of this work and our previous paper provides a relatively complete view of the pre-trained model sharing and distribution practices. 
Specifically, this paper discusses additional implications not covered in our previous work such as information transparency, quality indicators, model dependencies, release management, and security. 

\section{Threats to Validity}
\label{sec:threatToValidity}

\subsection{Construct Validity}

Although the information elements in RQ1 are extracted from multiple resources (like original webpages, definition files, JSON data structure), they may be incomplete. To mitigate this threat, the lead author and two of the other authors performed independent analysis of the studied package repositories to verify the list of identified IEs. 
We also verify the information elements with previous work~\cite{xiu2020exploratory}\cite{bommarito2019empirical}. 


\subsection{Internal Validity}

As discussed in the research questions, the release management processes for ML packages are not well-defined. In particular, some of the newer versions of ML packages are either shown within the same product pages (e.g., V1 to V4 of the same ML package are organized in the same 
page$\footnote{https://tfhub.dev/google/imagenet/mobilenet\_v1\_100\_128/feature\_vector/4}$) or on separate product pages (e.g., two versions of a ML package trained on different algorithms are displayed in different ML package pages$\footnote{https://tfhub.dev/google/imagenet/mobilenet\_v1\_100\_128/feature\_vector/4}^, $$\footnote{https://tfhub.dev/google/imagenet/mobilenet\_v2\_100\_128/feature\_vector/4}$, packages in the same algorithm but trained on different datasets and saved into different formats are also in separate pages$\footnote{https://tfhub.dev/google/bert\_multi\_cased\_L-12\_H-768\_A-12/1}^,$$\footnote{https://tfhub.dev/tensorflow/bert\_multi\_cased\_L-12\_H-768\_A-12/1}$).  
However, in software package repositories, new versions of a package will be generally organized in the same product page (e.g., all the versions of the TensorFlow, as Python package, are organized on the same page). 
When counting the number of ML packages, we actually count the number of separate ML package pages. These different criteria of counting the number of ML packages and packages may threaten the results of our repository comparisons. 

\subsection{External Validity}

ML package repositories are relatively new and they are rapidly changing. For example, 
we notice during our analysis that TFHub's information elements and JSON data structure changes over time (e.g., new information elements are added and the order of IE was changed). Hence, although our current findings are useful for practitioners and SE researchers, they may be out-dated in a few years. It would be worthwhile to replicate this study after a period of time to analyze the evolution of software practices (e.g., versioning, usage, package organization) within the ML package repositories. 

\section{Conclusions and Future Work}
\label{sec:conclusions}


This paper is an exploratory study on ML package repositories: TFHub and PyTorch Hub. 
First, we compare the information elements between ML package repositories and software package repositories, and then between two ML package repositories. We discovered some concerns of ML package repositories in terms of dependency management, release information, 
popularity indicator security and technical documentation information transparency. 
The second research question is about the contents (packages and models within) of ML package repositories. We looked into how the packages are classified and organized, the similarities between packages and models across different ML package repositories, and the release management practices of ML packages. 
The third research question is closely related to the basic functionalities (loading and using a package) and usage contexts the ML libraries support. We discovered that the ML package upgrade and removal functionalities are still missing in ML libraries. Also the signature and entrypoint mechanisms provided by TensorFlow and PyTorch, as well as the family phenomenon may bring about package evolution challenges.

Based on our findings, we recommend that 
(1) ML package repositories should provide information transparency by organizing ML related information elements in a structural manner. 
(2) ML libraries should support package upgrade and removal functionality. In addition, they should adopt similar release management practices from the domain of software package repositories. 
(3) PyTorch Hub packages should provide enough details for entrypoints in package documentation for ease of use by users. 
(4) Though the family phenomenon, signature mechanism and entrypoint mechanism are helpful for users, they should adopt product line architecture practices from traditional software engineering.

In the future, we will look into package evolution practices, like the detailed reasons why a new version is released and how APIs change.




\bibliographystyle{unsrt}
\bibliography{xmkBibLib}

\end{document}